\documentclass[aps,prl,twocolumn,superscriptaddress]{revtex4}%groupedaddress
\usepackage{graphicx}% Include figure files
\usepackage{dcolumn}% Align table columns on decimal point
\usepackage{bm}% bold math
\usepackage{amsmath}
\usepackage{amssymb}
\usepackage{mathrsfs}
\usepackage{float}
\usepackage{latexsym}
\usepackage{epsfig}
\usepackage{amsbsy}
\usepackage{array}
\usepackage{amssymb}
\usepackage{bm}
\usepackage{appendix} 
\usepackage{color}
\usepackage{tikz-feynman}
\usepackage{fancyhdr}
\usepackage{ulem}
\usepackage{cancel}
\usepackage{booktabs}
\usepackage{pifont}
\usetikzlibrary{calc}
\usepackage{tikz}
\usepackage{enumerate}
\usepackage[colorlinks=true,linkcolor=red,citecolor=blue,urlcolor=blue,]{hyperref}
\usepackage[doipre={doi:~}]{uri}
\usepackage{graphicx,scalerel}
\newcommand\sbullet[1][.5]{\mathbin{\ThisStyle{\vcenter{\hbox{%
  \scalebox{#1}{$\SavedStyle\bullet$}}}}}%
}

\begin{document}
\bibliographystyle{plain}
	
	\title{DC Current Generation and Power Feature in Strongly Driven Floquet-Bloch Systems}% 
	
	\author{Qiang Gao}
	\affiliation{Department of Physics, The University of Texas at Austin, Texas 78712, USA}
	
	\author{Yafei Ren}
	\affiliation{Department of Physics, The University of Texas at Austin, Texas 78712, USA}
	\affiliation{Department of Materials Science and Engineering, University of Washington, Seattle, Washington 98195, USA}

    \author{Qian Niu} 
	\affiliation{Department of Physics, The University of Texas at Austin, Texas 78712, USA}
    \affiliation{ICQD/HFNL and School of Physics, University of Science and Technology of China, Hefei, Anhui 230026, China}

	\date{\today}
	
	\begin{abstract}
	   We study the DC current generation in a periodically driven Bloch system connected to a heat bath. Under a relaxation time approximation, the density matrix for such a system is obtained, which is related to two equilibria: a Floquet quasi-equilibrium where the density matrix is diagonal under the Floquet-Bloch eigenbasis and an instantaneous Bloch thermal equilibrium. Then, the current responses and their power features, i.e. the power input behavior, are discussed in a unified manner, which reveals that there exist an intrinsic current and an extrinsic correction. Remarkably, the intrinsic part consumes no energy and corresponds to the Floquet quasi-equilibrium, while the extrinsic part needs a sustained energy input and originates from a shift between two equilibrium ensembles. We further investigate that role of the external driving field strength finding that large DC currents can be generated under a relatively strong but not too strong driving field.
	\end{abstract}

	\maketitle

\textit{Introduction}.---
Periodically driven systems or Floquet systems have long been an intensively studied area in physics, since they provide controllable platforms for realizing many exotic physical phenomena such as the Floquet topological insulator~\cite{rechtsman2013photonic,lindner2011floquet} and the (space-)time crystal~\cite{else2016floquet,zhang2017observation,autti2018observation,qiang2021floquet}. The external driving field has a wide range of possibilities: mechanical~\cite{wallquist2009hybrid,li2020enhancing}, optical~\cite{oka2019floquet,atanasov1996coherent}, or even acoustic~\cite{qiang2021floquet} drives, among which the optically driven systems have drawn the most attention due to their applications in industry like solar cell~\cite{nelson2003physics}. However, unlike the weakly driven system that can be dealt with using the perturbation theory~\cite{atanasov1996coherent}, for periodically driven systems with strong external field strength, Floquet analysis is necessary~\cite{ho1986floquet,kohler2005driven}. Recently, Floquet systems connected to heat sources characterizing more realistic open environments have also attracted many attentions~\cite{dehghani2014dissipative,dehghani2015out,seetharam2015controlled,iadecola2015floquet,Shirai2015Condition,sato2020floquet}. Being essentially energy non-conserved, Floquet systems constantly involve energy conversions between different forms~\cite{qiang2021floquet,sato2020floquet}, whose power features are extremely important for energy conversion applications~\cite{bai2020floquet}.

In this work, we consider a system with periodicity in both space and time, known as the Floquet-Bloch system, connected to a heat bath, and investigate the DC current generations and their power features. We first characterize the heat bath through a relaxation time approximation~\cite{ho1986floquet,sato2019microscopic}, and obtain the density matrix for a non-interacting Floquet-Bloch system with damping, which is related to two equilibria: an instantaneous Bloch thermal equilibrium and a Floquet quasi-equilibrium whose density matrix is diagonal under the Floquet-Bloch eigenbasis. In the weakly damping but strongly driven regime, we find that the current responses consists of an intrinsic part and an extrinsic correction due to the damping, which are nonzero for systems with broken time-reversal and inversion symmetries. The power features are the following: the intrinsic part is dissipationless and related to the Floquet quasi-equilibrium, while the extrinsic part needs a sustained energy input and originates from a shift between two equilibrium ensembles. Thus the generation of the intrinsic current enhances the overall power efficiency. We also investigate the role of the driving field strength in the DC current generations revealing that the currents become larger under a relatively stronger field but drops dramatically at a certain point where a Floquet gap transition happens.

\textit{Floquet Analysis}.---
We start from a non-driven Bloch system characterized by: $\hat{H}_B(\boldsymbol{r}) |\varphi^\alpha_{\boldsymbol{k}}(\boldsymbol{r})\rangle = E_\alpha (\boldsymbol{k})|\varphi^\alpha_{\boldsymbol{k}}(\boldsymbol{r})\rangle$,
where $\hat{H}_B(\boldsymbol{r})=\hat{H}_B(\boldsymbol{r}+\boldsymbol{R})$ is the static Hamiltonian with $N_B$ multiple bands, $\alpha\in\{1,2,\cdots,N_B\}$ stands for the Bloch band index, and $\boldsymbol{k}$ is the lattice momentum. According to Bloch's Theorem, the wavefunction can be expressed as a plane wave multiplied by a spatially periodic function: $|\varphi^\alpha_{\boldsymbol{k}}(\boldsymbol{r})\rangle = e^{i\boldsymbol{k}\cdot\boldsymbol{r}}|u^\alpha_{\boldsymbol{k}}(\boldsymbol{r})\rangle$, where $|u^\alpha_{\boldsymbol{k}}(\boldsymbol{r}+\boldsymbol{R})\rangle = |u^\alpha_{\boldsymbol{k}}(\boldsymbol{r})\rangle$. Then, we apply the time-periodic driving field $\boldsymbol{A}(t)$ to the Bloch system which makes $\hat{H}_B \to \hat{H}(\boldsymbol{r},t)$ with periodicity in time $\hat{H}(\boldsymbol{r},t) = \hat{H}(\boldsymbol{r},t+T)$, and the Schr\"odinger equation reads
\begin{equation}\label{timeSE}
    \hat{H}(\boldsymbol{r},t)|\Psi(\boldsymbol{r},t)\rangle = i\hbar\partial_t |\Psi(\boldsymbol{r},t)\rangle.
\end{equation}

To solve~Eq.\eqref{timeSE}, we introduce an extended Floquet-Bloch basis $\{|\phi^\alpha_{n,\boldsymbol{k}}(\boldsymbol{r},t)\rangle \equiv e^{-in\Omega t}|\varphi^\alpha_{\boldsymbol{k}}(\boldsymbol{r})\rangle\}$ which is the original Bloch wavefunction shifted in energy by integer multiples of $\Omega\equiv 2\pi/T$~\cite{qiang2021floquet,gomez2013floquet}. Here $n$ is the Floquet multiplicity. Under the extended Floquet-Bloch basis, we can expand the wavefunction as
\begin{equation}\label{FBeigenstate}
\begin{split}
     |\Psi^\mu_{\boldsymbol{k}}(\boldsymbol{r},t)\rangle =e^{-i\omega_\mu(\boldsymbol{k})t} \sum_{n,\beta}f^{\mu,\beta}_{n,\boldsymbol{k}}|\phi^\beta_{n,\boldsymbol{k}}(\boldsymbol{r},t)\rangle
\end{split}
\end{equation}
where $\omega_\mu$ is the quasi-energy with band index $\mu$. We then denote
$\sum_{n,\beta}f^{\mu,\beta}_{n,\boldsymbol{k}}|\phi^\beta_{n,\boldsymbol{k}}(\boldsymbol{r},t)\rangle \equiv e^{i\boldsymbol{k}\cdot\boldsymbol{r}}|\Tilde{u}_{\boldsymbol{k}}^\mu(\boldsymbol{r},t)\rangle$
as the corresponding Floquet-Bloch eigenstate with $|\Tilde{u}_{\boldsymbol{k}}^\mu(\boldsymbol{r},t)\rangle$ referring to the Floquet-Bloch periodic function that has the same periodicity as the Hamiltonian~\cite{Notations}. The expansion coefficient $f^{\mu,\beta}_{n,\boldsymbol{k}}$ satisfies the following eigen-equation: $\sum_{n,\alpha}\mathcal{H}_{m,n;\beta,\alpha}f^{\mu,\alpha}_{n,\boldsymbol{k}}=\hbar\omega_\mu f^{\mu,\beta}_{
m,\boldsymbol{k}}$ with the kernel matrix
\begin{equation}\label{HamiltonianKernel}
    \begin{split}
    \mathcal{H}_{m,n;\alpha,\beta} &\equiv \langle\langle \phi^\alpha_{m,\boldsymbol{k}}(\boldsymbol{r},t) | \hat{H}(\boldsymbol{r},t)-i\hbar\partial_t|\phi^\beta_{n,\boldsymbol{k}}(\boldsymbol{r},t)\rangle\rangle,
    \end{split}
\end{equation}
where $\langle\langle\cdot\rangle\rangle \equiv 1/T\int_0^T \langle\cdot\rangle$ is the spacetime inner product.
The quasi-energy $\omega_\mu$ inherits the degrees of freedom of the Bloch system, so in principle, we will have $N_B$ different quasi-energy bands $\{\omega_1,\omega_2,\cdots,\omega_{N_B}\}$ within a Floquet Brillouin zone. More details about the matrix $\mathcal{H}_{m,n;\alpha,\beta}$ are given in the Supplementary Material~\cite{SM}.

\textit{Approach to Floquet Quasi-Equilibrium}.--- We now consider a Floquet-Bloch system connected to a heat bath which introduces damping. Such damping system is governed by the following quantum Liouville equation ($\hbar = 1$)~\cite{ho1986floquet,sato2019microscopic}:
\begin{equation}\label{QLE}
    i\partial_t \hat{\rho} (\boldsymbol{k},t) = [\hat{H}(\boldsymbol{k},t), \hat{\rho} (\boldsymbol{k},t)] + i[\hat{D},\hat{\rho}(\boldsymbol{k},t)],
\end{equation}
where $\hat{\rho}(\boldsymbol{k},t)$ is the density operator of the system at fixed $\boldsymbol{k}$ and $\hat{H}(\boldsymbol{k},t)\equiv e^{-i\boldsymbol{k}\cdot\boldsymbol{r}}\hat{H}(\boldsymbol{r},t)e^{i\boldsymbol{k}\cdot\boldsymbol{r}}$ is the Hamiltonian of the corresponding non-damping system. In this work, we will restrict our discussion in $\boldsymbol{k}$-conserved systems. The density matrix operator can be expanded using either the static Bloch eigenbasis $\{|u^\alpha_{\boldsymbol{k}}\rangle\}$ or the Floquet-Bloch eigenbasis $\{|\Tilde{u}^\mu_{\boldsymbol{k}}(t)\rangle\}$ (index $\boldsymbol{r}$ is omitted):
\begin{equation}\label{densityMatrix}
\begin{split}
   \hat{\rho} (\boldsymbol{k},t) &=\sum_{\alpha,\beta}\rho^B_{\alpha,\beta}(\boldsymbol{k},t)|u^\alpha_{\boldsymbol{k}}\rangle\langle u^\beta_{\boldsymbol{k}}| \\
    &= \sum_{\mu,\nu}\rho^F_{\mu,\nu}(\boldsymbol{k},t)|\Tilde{u}^\mu_{\boldsymbol{k}}(t)\rangle\langle\Tilde{u}^\nu_{\boldsymbol{k}}(t)|,
\end{split}
\end{equation}
where $\rho^B_{\alpha,\beta}(\boldsymbol{k},t)$ and $\rho^F_{\mu,\nu}(\boldsymbol{k},t)$ are the density matrix elements under two eigenbases, respectively. The operator $\hat{D}$ characterizes the damping due to the heat transfer and the electron scattering. The role of damping is to relax the system to a thermal equilibrium with the heat bath, while the Floquet drive is to strike the system out of equilibrium.
Under the relaxation time approximation~\cite{sato2019microscopic,relaxationTimeApproximation}, we can write
\begin{equation}\label{dampingTerm}
    \begin{split}
    [\hat{D},\hat{\rho}(\boldsymbol{k},t)] = &-\Gamma[\hat{\rho} (\boldsymbol{k},t)-\hat{\rho}^{B,eq} (\boldsymbol{k},t)] \\
    &-\sum_{\alpha\ne \beta}\Gamma'_{\alpha,\beta}(\boldsymbol{k})\rho^B_{\alpha,\beta}(\boldsymbol{k},t)|u^\alpha_{\boldsymbol{k}}\rangle\langle u^\beta_{\boldsymbol{k}}|
    \end{split}
\end{equation}
% \begin{equation}\label{dampingTerm}
%     \begin{split}
%         &[\hat{D},\hat{\rho}(\boldsymbol{k},t)]_{\alpha,\alpha} = -\Gamma\left[\rho^B_{\alpha,\alpha}(\boldsymbol{k},t)-\rho^{B,eq}_{\alpha,\alpha}(\boldsymbol{k}\textcolor{blue}{,t})\right]|u^\alpha_{\boldsymbol{k}}\rangle\langle u^\alpha_{\boldsymbol{k}}| ,\\
%         &[\hat{D},\hat{\rho}(\boldsymbol{k},t)]_{\alpha,\beta} = -\Gamma_{\alpha,\beta}(\boldsymbol{k})\rho^B_{\alpha,\beta}(\boldsymbol{k},t)|u^\alpha_{\boldsymbol{k}}\rangle\langle u^\beta_{\boldsymbol{k}}| ,
%     \end{split}
% \end{equation}
where the first line characterizes the thermal contact with damping rate $\Gamma$, and the second line is the pure dephasing caused by the electron scattering~\cite{ho1986floquet} with dephasing rate $\Gamma'_{\alpha,\beta}$.
The $\hat{\rho}^{B,eq} (\boldsymbol{k},t)$ is an instantaneous Bloch equilibrium, which is not diagonal under the static Bloch eigenbasis: $\hat{\rho}^{B,eq} (\boldsymbol{k},t)=\sum_{\alpha,\beta}\rho^{B,eq}_{\alpha,\beta}(\boldsymbol{k},t)|u^\alpha_{\boldsymbol{k}}\rangle\langle u^\beta_{\boldsymbol{k}}|$~\cite{SM}. This equilibrium indicates a field dragging effect (see Eq.~\eqref{ins_Blo_den} for a special case).

Under the Floquet-Bloch eigenbasis, we can rewrite the quantum Liouville equation as~\cite{SM}:
\begin{equation}\label{QLEinFlouqet}
    i\partial_t\rho^F_{\mu,\nu} = \left[ \omega_\mu(\boldsymbol{k})- \omega_\nu(\boldsymbol{k})-i\Gamma\right]\rho^F_{\mu,\nu}+i\Gamma\Bar{\rho}^{eq}_{\mu,\nu}-i\mathcal{S}_{\mu,\nu},
\end{equation}
where $\Bar{\rho}^{eq}_{\mu,\nu}(\boldsymbol{k},t)\equiv\sum_{\alpha,\beta}\rho^{B,eq}_{\alpha,\beta}(\boldsymbol{k},t)\langle\Tilde{u}^\mu_{\boldsymbol{k}}(t) |u^\alpha_{\boldsymbol{k}}\rangle\langle u^\beta_{\boldsymbol{k}}|\Tilde{u}^\nu_{\boldsymbol{k}}(t)\rangle$ is still the instantaneous Bloch equilibrium but expanded in the Floquet eigenbasis, and $\mathcal{S}_{\mu,\nu}(\boldsymbol{k},t)\equiv\sum_{\alpha\ne\beta}\Gamma'_{\alpha,\beta}(\boldsymbol{k})\langle\Tilde{u}^\mu_{\boldsymbol{k}}(t) |\rho^B_{\alpha,\beta}(\boldsymbol{k})|u^\alpha_{\boldsymbol{k}}\rangle\langle u^\beta_{\boldsymbol{k}}|\Tilde{u}^\nu_{\boldsymbol{k}}(t)\rangle$ is the scattering contribution. It is hard to solve the differential equation above, especially when the scattering matrix $\mathcal{S}_{\mu,\nu}$ is unknown. However, the system will evolve into a steady state similar to a forced damping oscillator whose changing frequency synchronizes with the external driving frequency. Therefore, we can do the Fourier transformations: $\rho^F_{\mu,\nu}(\boldsymbol{k},t) = \sum_l \rho^{F,l}_{\mu,\nu}(\boldsymbol{k})e^{-il\Omega t}$, $\Bar{\rho}^{eq}_{\mu,\nu}(\boldsymbol{k},t) = \sum_l \Bar{\rho}^{eq,l}_{\mu,\nu}(\boldsymbol{k})e^{-il\Omega t}$, and $\mathcal{S}_{\mu,\nu}(\boldsymbol{k},t) = \sum_l \mathcal{S}^{l}_{\mu,\nu}(\boldsymbol{k})e^{-il\Omega t}$.
By inserting those Fourier expansions into Eq.~\eqref{QLEinFlouqet}, we obtain a solution for the density matrix:
\begin{equation}\label{FourierComponent}
    \rho^{F,l}_{\mu,\nu}(\boldsymbol{k}) = \frac{-i\Gamma\Bar{\rho}^{eq,l}_{\mu,\nu}(\boldsymbol{k})+i\mathcal{S}^{l}_{\mu,\nu}(\boldsymbol{k})}{\omega_\mu(\boldsymbol{k})-\omega_\nu(\boldsymbol{k})-l\Omega-i\Gamma}.
\end{equation}
% We note that the matrix element $\mathcal{S}^{l}_{\mu,\nu}(\boldsymbol{k})$ depends on the pure dephasing rate $\Gamma'_{\alpha,\beta}$ and the density matrix $\rho^{F,l}_{\mu,\nu}(\boldsymbol{k})$, which requires the details of all possible interactions. 

To further simplify our considerations, we introduce two extremes: one is the pure thermal contact, $\Delta_{\boldsymbol{A}},\Gamma\gg\Gamma'_{\alpha,\beta}$, where the electron-electron interaction is surpassed, for example, by large dielectric constant of the material; the other is the isolated system, $\Delta_{\boldsymbol{A}},\Gamma'_{\alpha,\beta}\gg\Gamma$, where the system is isolated from any heat sources. The $\Delta_{\boldsymbol{A}}=\min_{\mu\ne\nu,l}\{ |\omega_\mu(\boldsymbol{k})-\omega_\nu(\boldsymbol{k})-l\Omega|\}$ is the minimum direct gap of the Floquet-Bloch system. Our main interest in this work is the first extreme where we can approximately let $\Gamma'_{\alpha,\beta} = 0$, corresponding to a non-interacting Floquet-Bloch system. Thus, the density matrix becomes:
\begin{equation}\label{FBdensityMatrix}
\begin{split}
    &\rho^F_{\mu,\nu}(\boldsymbol{k},t;\Gamma) = \\
    &\sum_{n,m,l,\alpha,\beta}-i\Gamma e^{-il\Omega t}\frac{\rho^{B,eq,m}_{\alpha,\beta}(\boldsymbol{k})\left(f_{n,\boldsymbol{k}}^{\mu,\alpha}\right)^*f_{n+l-m,\boldsymbol{k}}^{\nu,\beta}}{\omega_\mu(\boldsymbol{k})-\omega_\nu(\boldsymbol{k})-l\Omega-i\Gamma},
\end{split}
\end{equation} 
which is in general non-diagonal and time-dependent. Here $ \rho^{B,eq,m}_{\alpha,\beta}(\boldsymbol{k})$ is the Fourier component of $\rho^{B,eq}_{\alpha,\beta}(\boldsymbol{k},t) = \sum_{m}\rho^{B,eq,m}_{\alpha,\beta}(\boldsymbol{k})e^{-im\Omega t}$. One can check that this density matrix is Hermitian and more importantly gauge invariant under the gauge transformation $|u^\alpha_{\boldsymbol{k}}\rangle\to e^{-i\theta^\alpha_{\boldsymbol{k}}}|u^\alpha_{\boldsymbol{k}}\rangle$.

The density matrix in Eq.~\eqref{FBdensityMatrix} indicates two special $\Gamma$-independent ensembles. 
Either, through a very large damping ($\Gamma\to \infty$), the system is quickly relaxed to an instantaneous Bloch equilibrium with the heat bath, in which the density matrix is reduced to $\hat{\rho}^{B,eq}(\boldsymbol{k},t)$.
% \begin{equation}\label{equilibria}
%     \hat{\rho}^{B,eq}(\boldsymbol{k},t) = \sum_{\alpha}\rho^{B,eq}_{\alpha,\alpha}(\boldsymbol{k}\textcolor{blue}{,t})|u^\alpha_{\boldsymbol{k}}(\boldsymbol{r})\rangle\langle u^\alpha_{\boldsymbol{k}}(\boldsymbol{r})|.
% \end{equation}
Or, by turning off the thermal contact ($\Gamma\rightarrow 0$), we obtain a diagonal density matrix: 
\begin{equation}\label{densityMatrixWithoutGamma}
    \rho^{F,eq}_{\mu,\nu}(\boldsymbol{k}) = \delta_{\mu,\nu}\sum_{n,m,\alpha,\beta}\rho^{B,eq,m}_{\alpha,\beta}(\boldsymbol{k})\left(f_{n,\boldsymbol{k}}^{\mu,\alpha}\right)^*f_{n-m,\boldsymbol{k}}^{\nu,\beta},
\end{equation}
where we have used the fact that $\lim_{\Gamma\rightarrow 0}\frac{-i\Gamma}{\omega_\mu - \omega_\nu - l\Omega-i\Gamma} = \delta_{\mu,\nu}\delta_{l,0}$. One can also show that this is the instantaneous Bloch equilibrium projected onto the Floquet-Bloch eigenbasis with off-diagonal terms erased by damping: $\rho^{F,eq}_{\mu,\mu}=\langle\langle\tilde{u}^\mu_{\boldsymbol{k}}(t)|\hat{\rho}^{B,eq}(\boldsymbol{k},t)|\tilde{u}^\mu_{\boldsymbol{k}}(t)\rangle\rangle$, which indicates that with no heat contact (i.e. isolated system), it is a so-called diagonal ensemble~\cite{Rigol2008Thermalization,d2014long}:
\begin{equation}\label{equilibria}
    \hat{\rho}^{F,eq}(\boldsymbol{k},t)= \sum_{\mu}\rho^{F,eq}_{\mu,\mu}(\boldsymbol{k})|\tilde{u}^\mu_{\boldsymbol{k}}(t)\rangle\langle\tilde{u}^\mu_{\boldsymbol{k}}(t)|.
\end{equation}
In the remainder, we refer to this diagonal ensemble as the Floquet quasi-equilibrium since its density matrix element is diagonal and time-independent under the Floquet-Bloch eigenbasis.

\textit{Floquet-Liouville Equation: DC Current Response and Power Input}.--- The essence of a Floquet system is that it evolves periodically in time, which superficially renders the fact that for an operator $\hat{\mathcal{O}}$ also periodic in time, we have $\int_{0}^{T}\frac{d}{dt}\text{tr}(\hat{\rho}\hat{\mathcal{O}})dt =\text{tr}(\hat{\rho}\hat{\mathcal{O}})|_{t=T}-\text{tr}(\hat{\rho}\hat{\mathcal{O}})|_{t=0}= 0$. One can combine such fact with the quantum Liouville equation to obtain the following Floquet-Liouville equation under the Floquet-Bloch eigenbasis~\cite{SM}:
\begin{equation}\label{FLe}
    \text{Tr}^F[\hat{\rho}\partial_t\hat{\mathcal{O}}]=-i\text{Tr}^F[\hat{\rho}[\hat{H}-\omega,\hat{\mathcal{O}}]]-\text{Tr}^F[[\hat{D},\hat{\rho}]\hat{\mathcal{O}}],
\end{equation}
where $\text{Tr}^F[\sbullet]\equiv\sum_{\mu}\langle\langle\Tilde{u}^\mu_{\boldsymbol{k}} (t)|\sbullet|\Tilde{u}^\mu_{\boldsymbol{k}}(t)\rangle\rangle$ is the Floquet spacetime trace and $\omega$ is the quasi-energy of corresponding Floquet eigenstate: $\omega|\Tilde{u}^\mu_{\boldsymbol{k}}(t)\rangle\rightarrow\omega_\mu(\boldsymbol{k})|\Tilde{u}^\mu_{\boldsymbol{k}}(t)\rangle$. This Floquet-Liouville equation allows us to study many DC responses in periodically driven systems in a unified manner. In this section, we focus on the DC current response and the driven power. The discussion will be based on the main result in Eq.~\eqref{FBdensityMatrix} and the assumptions for obtaining it, where we ignored the electron-electron interactions. In such setting, the damping term in Eq.~\eqref{dampingTerm} can be reduced to $[\hat{D},\hat{\rho}]=-\Gamma(\hat{\rho}-\hat{\rho}^{B,eq}) $.

The DC charge current can be calculated by (ignoring $-e$ from now on): $ \boldsymbol{j}_{DC}  = \int\text{d}\boldsymbol{k}\text{Tr}^F[\hat{\rho}\hat{\boldsymbol{v}}]$ with $ \hat{\boldsymbol{v}}\equiv\partial_{\boldsymbol{k}}\hat{H}(\boldsymbol{k},t)$ the velocity operator. In favor of Eq.~\eqref{FLe}, we choose $\hat{\mathcal{O}}=i\partial_{\boldsymbol{k}}$ and end up with the following result~\cite{SM}:
\begin{equation}\label{velocityEV}
    \text{Tr}^F[\hat{\rho}\hat{\boldsymbol{v}}] = \text{Tr}^F[\hat{\rho}^{F,eq}\hat{\boldsymbol{v}}] +  \Gamma\times\text{Tr}^F[(\hat{\rho}^{F,eq}-\hat{\rho}^{B,eq})i\partial_{\boldsymbol{k}}]+\Lambda,
\end{equation}
where $\Lambda=O(\frac{\Gamma^2}{\Delta_{\boldsymbol{A}}-i\Gamma})$ is a correction that contains higher orders of $\Gamma$ and is negligible when $\Gamma\ll \Delta_{\boldsymbol{A}}$. One can treat $\text{Tr}^F[\hat{\rho}^{F(B),eq}i\partial_{\boldsymbol{k}}]$, the ensemble average of the operator $i\partial_{\boldsymbol{k}}$, as the polarization projected onto the Floquet (Bloch) eigenspace. In this work, we focus on the weakly damping but strongly driven regime where $\Lambda\to 0$.

We want to make the following remarks for the result in the Eq.~\eqref{velocityEV}: (i) it is gauge invariant~\cite{SM}, allowing us to choose any gauge for convenience; (ii) the linear order correction in $\Gamma$ is geometrical and related to a shift in polarization between two (quasi-)equilibrium ensembles; (iii) via a gauge transformation: $|\tilde{u}^\mu_{\boldsymbol{k}}(t)\rangle\to e^{-i\theta^\mu_{\boldsymbol{k}}}|\tilde{u}^\mu_{\boldsymbol{k}}(t)\rangle$ with $\theta^\mu_{\boldsymbol{k}} = \int_{\boldsymbol{k}_0}^{\boldsymbol{k}}d\boldsymbol{k}' \langle\langle\tilde{u}^\mu_{\boldsymbol{k}'}(t) |\hat{\rho}^{B,eq}i\partial_{\boldsymbol{k}'}|\tilde{u}^\mu_{\boldsymbol{k}'}(t)\rangle\rangle/\rho^{F,eq}_{\mu,\mu}(\boldsymbol{k}')$, we can rewrite Eq.~\eqref{velocityEV} as $\text{Tr}^F[\hat{\rho}\hat{\boldsymbol{v}}] = \text{Tr}^F[\hat{\rho}^{F,eq}\Tilde{\hat{\boldsymbol{v}}}]$ with $\Tilde{\hat{\boldsymbol{v}}}\equiv\hat{\boldsymbol{v}}+\Gamma i\partial_{\boldsymbol{k}}$ defined as the shifted velocity operator. Then the current reads:
\begin{equation}\label{currentInFloquet}
    \boldsymbol{j}_{DC}=\int \text{d}\boldsymbol{k}\sum_{\mu}\rho^{F,eq}_{\mu,\mu}\left(\frac{\partial\omega_{\mu}}{\partial\boldsymbol{k}}+\Gamma\boldsymbol{\mathcal{A}}^F_{\mu}\right)\equiv\boldsymbol{j}_{DC}^{\text{In}} + \boldsymbol{j}_{DC}^{\text{Ex}},
\end{equation}
where $\boldsymbol{\mathcal{A}}^F_{\mu}\equiv\langle\langle\tilde{u}^\mu_{\boldsymbol{k}}(t) |i\partial_{\boldsymbol{k}}|\tilde{u}^\mu_{\boldsymbol{k}}(t)\rangle\rangle $ is the Floquet Berry connection after choosing the above gauge~\cite{GaugeFixing}. The current is now fully diagonal in the Floquet band index $\mu$ and contains: $\boldsymbol{j}_{DC}^{\text{In}}$, $\Gamma$-independent and thus an intrinsic current; $ \boldsymbol{j}_{DC}^{\text{Ex}}$, linear in $\Gamma$ and an extrinsic contribution that reflects a shift in polarization between two ensembles. It is worth noting that the current response written in Eq.~\eqref{currentInFloquet} manifests its possible topological natures as the Berry connection comes into play~\cite{Morimoto2016Topological}.

There are requirements for having nonzero current responses. The first requirement is about symmetry. The extrinsic current $\boldsymbol{j}_{DC}^{\text{Ex}}$ is related to a shift in the polarization of electrons, which is nonzero if the inversion symmetry (parity) is broken. However, for the intrinsic current $\boldsymbol{j}_{DC}^{\text{In}}$, we also need to break the time-reversal symmetry (TRS), since $\partial_{\boldsymbol{k}}\omega_{\mu}$ is odd under TRS. Another requirement is special for $\boldsymbol{j}_{DC}^{\text{In}}$ that the quasi-equilibrium distribution $\rho^{F,eq}_{\mu,\mu}$ should not be a function of quasi-energy $\omega_{\mu}$, otherwise, the intrinsic current will always be zero due to the periodicity of $\omega_{\mu}$ in the Brillouin zone.

To generate the currents, we need to know how much energy should the Floquet drive provide to sustain the DC currents, i.e. the power features. We now specify the Floquet drive as coherent light where the Hamiltonian takes the form $ \hat{H}(\boldsymbol{k},t)=\hat{H}_B(\boldsymbol{k}+\boldsymbol{A}(t))$ with $\boldsymbol{A}(t)$ the vector potential. Then, the net power provided by the light over one Floquet period is $\Bar{P}_{\text{net}} =\int d\boldsymbol{k}\Bar{P}(\boldsymbol{k}) =\int d\boldsymbol{k}\text{Tr}^{F}[\boldsymbol{E}(t)\cdot\hat{\rho}\hat{\boldsymbol{v}}]$ with $ \boldsymbol{E}(t)=-\partial_t \boldsymbol{A}(t)$ the AC electric field. By choosing $\hat{\mathcal{O}}= \hat{H}(\boldsymbol{k},t)$ in Eq.~\eqref{FLe}, we get the input power at specific $\boldsymbol{k}$~\cite{SM}:
\begin{equation}
    \Bar{P}(\boldsymbol{k}) = \Gamma\times\text{Tr}^F[(\hat{\rho}^{F,eq}-\hat{\rho}^{B,eq})\hat{H}]+O(\frac{\Gamma^2}{\Delta_{\boldsymbol{A}}-i\Gamma}),
\end{equation}
indicating that the intrinsic current is dissipationless, while the extrinsic current needs a sustained power input. Such result is understandable since for intrinsic current, we ignored the interactions among electrons giving no internal damping channels; but for extrinsic current, damping is involved and the power is related to a energy shift between two ensembles.

It's important to see the power feature in energy non-conserved systems since it tells us the direction of energy flow which in our case is from Floquet drive to the heat bath. More remarkable is that intrinsic current is dissipationless which can be favored in energy conversion applications. Combined with the symmetry argument, breaking TRS would help to obtain such desirable current. As for the extrinsic current, it becomes dominant in TRS preserved systems and more importantly its power efficiency defined as $\boldsymbol{j}_{DC}^{\text{Ex}}/\Bar{P}_{\text{net}}$ reflects some band properties~\cite{SM}. The field dependence of the DC currents and power feature is also important and will be discussed in the next section.

% In real engineering, damping is always present and the TRS is normally preserved, thus the goal is to maximize the extrinsic current while minimize the damping. To further characterize such optimization, we define a conversion efficiency as the ratio between the extrinsic DC current and the input AC power: $F^\text{Ex}=|\boldsymbol{j}_{DC}^{\text{Ex}}|/\Bar{P}_{net}$. 

\begin{figure}
    \centering
    \includegraphics[width=8.5cm]{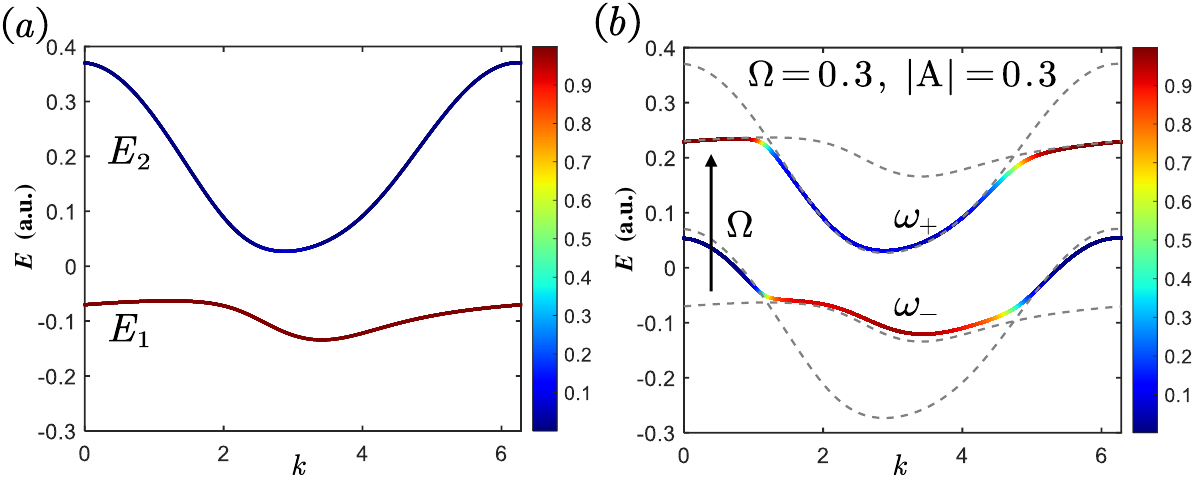}
    \caption{(a) The thermal distribution of electrons sitting on a two-band Bloch system described by Eq.~\eqref{ModelHamiltonian} with $\varepsilon_1 = 0.1$, $\varepsilon_2 = 0$, $t_1 = 0.11$, $t_2 = 0.05+0.02i$, $t_3 = 0.1$, $t_4 = 0$, $d = 1$, and $s = 0.2$. The chemical potential and temperature are set to be $\mu_c = 0$ and $k_B\mathcal{T}=0.01 $, respectively. (b) The Floquet quasi-equilibrium distribution according to Eq.~\eqref{densityMatrixWithoutGamma}, after turning on the Floquet drive with frequency $\Omega=0.3$ and amplitude $|\boldsymbol{A}|=0.3$. The distributions are reflected by colors with color bars alongside each panel.}
    \label{fig:FBbands}
\end{figure}

\begin{figure}
    \centering
    \includegraphics[width=8.5cm]{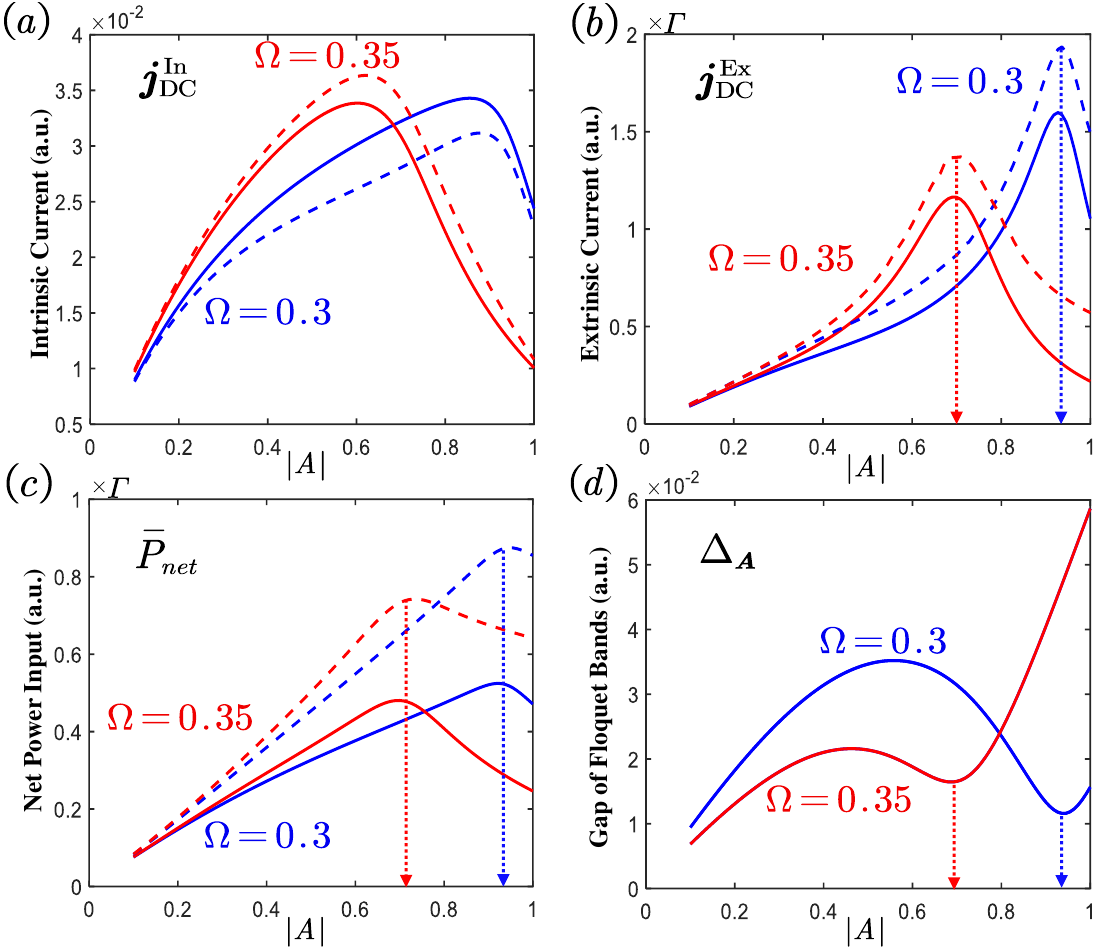}
    \caption{(a) The intrinsic current $\boldsymbol{j}_{DC}^{\text{In}}$, (b) the extrinsic current $\boldsymbol{j}_{DC}^{\text{Ex}}$, (c) the net power input $ \Bar{P}_{\text{net}}$, and (d) the Floquet direct gap $\Delta_{\boldsymbol{A}}$ as functions of $|\boldsymbol{A}|$ for $\Omega=0.3$ (blue) and $\Omega=0.35$ (red), with (dashed) and without (solid) dragging effect included, respectively. We mark the turning points of curves in panels (b-d) using dotted arrows. In this calculation, we require that $\Gamma \ll \Delta_{\boldsymbol{A}}\sim 0.01$. Parameters used are the same as in Fig.~\ref{fig:FBbands}.}
    \label{fig:FR}
\end{figure}

\textit{Case Study: A Two-band System Connected to A Heat Bath}.--- Consider a two-band model in which we explicitly break the TRS and parity to see the DC current generations and power features under strong driving field. The setup is a two-band Bloch system $H_B(\boldsymbol{k})$ with $E_1(\boldsymbol{k})$, $E_2(\boldsymbol{k})$ and $|u^1_{\boldsymbol{k}}\rangle$, $|u^2_{\boldsymbol{k}}\rangle$ as its two gaped energy bands and corresponding eigenstates. Then a coherent linearly polarized light, $\boldsymbol{A}(t)=\boldsymbol{A}e^{i\Omega t}+\boldsymbol{A}^*e^{-i\Omega t}$, is applied to the Bloch system as a Floquet drive. The Hamiltonian then becomes $H(\boldsymbol{k},t)= H_B(\boldsymbol{k}+\boldsymbol{A}(t))$.
The system is then connected to a heat bath at temperature $\mathcal{T}$, which, at thermal equilibrium, obeys the Fermi-Dirac distribution: $\rho^{B,eq,s}_{\alpha,\alpha}(\boldsymbol{k}) =\frac{1}{e^{[E_{\alpha}(\boldsymbol{k})-\mu_c]/k_B \mathcal{T}}+1}$, with $\mu_c$ the chemical potential sitting in the gap. The instantaneous Bloch thermal equilibrium can be approximated as~\cite{SM}
\begin{equation}\label{ins_Blo_den}
    \rho^{B,eq}_{\alpha,\beta}(\boldsymbol{k},t) \approx \rho^{B,eq,s}_{\alpha,\alpha}\delta_{\alpha,\beta} + i\boldsymbol{A}(t)\cdot\boldsymbol{\mathcal{A}}^{B}_{\alpha,\beta}(\boldsymbol{k})(\rho^{B,eq,s}_{\alpha,\alpha}-\rho^{B,eq,s}_{\beta,\beta}),
\end{equation}
where $\boldsymbol{\mathcal{A}}^{B}_{\alpha,\beta}\equiv\langle u^{\alpha}_{\boldsymbol{k}}|i\partial_{\boldsymbol{k}}|u^{\beta}_{\boldsymbol{k}}\rangle$ is the Bloch berry connection, and the second term on the right hand side represents a field dragging effect.
The two Floquet bands denoted as $\omega_{\pm}(\boldsymbol{k})$ can be found by diagonalizing the corresponding kernel matrix in Eq.~\eqref{HamiltonianKernel}. One last thing is that the damping rate $\Gamma$ can in general depend on field and momentum, which makes the current and the total power hard to be evaluated. Here for simplicity, we assume a constant $\Gamma$. 

In Fig.~\ref{fig:FBbands} and Fig.~\ref{fig:FR}, we show some numerical results. The Bloch model is a 1D diatomic chain with Hamiltonian~\cite{BlochBandParameters}
\begin{equation}\label{ModelHamiltonian}
		H_B(k) =  \begin{bmatrix}
				\varepsilon_1 + 2t_3\cos(kd) & e^{-iks}(t_1+t_2 e^{ikd})  \\
	        	e^{iks}(t_1^*+t_2^* e^{-ikd}) & \varepsilon_2 + 2t_4\cos(kd)
			\end{bmatrix}
\end{equation}
and other set-ups  are the same as mentioned before, except for restricting in 1D. First, we plot in Fig.~\ref{fig:FBbands}(a) the static Bloch thermal equilibrium distribution and in Fig.~\ref{fig:FBbands}(b) the corresponding Floquet quasi-equilibrium distribution. Note that in the Bloch thermal equilibrium, the electron distribution is a function of the energy, while in the Floquet quasi-equilibrium, the distribution is not a function of the quasi-energy, as one can see from Fig.~\ref{fig:FBbands}(b) that $k$ points with same quasi-energies have different distributions.
The reason is that the Floquet quasi-equilibrium is indeed a non-equilibrium state. 

Notice that the 1D model with the parameters we choose largely breaks the TRS and parity, which induces nonzero current responses.
In Fig.~\ref{fig:FR}(a,b), we plot the intrinsic and extrinsic currents as functions of the field strength $|\boldsymbol{A}|$ for two different driven frequencies. Meanwhile, we also explicitly illustrate the influence of the dragging effect which basically gives numerical corrections. An important observation is that at relatively small $|\boldsymbol{A}|$ (but still $\Delta_{\boldsymbol{A}}\gg \Gamma$), we have approximately $\boldsymbol{j}_{DC}^{\text{In}}\propto |\boldsymbol{A}|$ and $\boldsymbol{j}_{DC}^{\text{Ex}}\propto \Gamma |\boldsymbol{A}|$, latter of which is consistent with the prediction made in Ref.~\cite{Morimoto2016Topological}. It is interesting to see that the net power input depicted in Fig.~\ref{fig:FR}(c) shows a strong linear dependence in field strength in a larger field regime. Such linearity is persistent even after a large correction due to the dragging effect.
However, when $|\boldsymbol{A}|$ is getting larger, the complexity appears that the currents and power input start to decrease significantly at some turning points. This feature then tells us that there exists an optimized field strength for current generation, namely the DC currents should be generated with strong but not too strong field. Here we give a phenomenological explanation that the decreases in those quantities are consequences of large band shifting under strong driving. To confirm this, we plot the Floquet direct band gaps in Fig.~\ref{fig:FR}(d). When $|\boldsymbol{A}|$ is fairly small, the gap opens at the resonant points where $E_1+\Omega-E_2 = 0$ (see Fig.~\ref{fig:FBbands}(b)), and grows linearly in $|\boldsymbol{A}|$. However, as $|\boldsymbol{A}|$ gets larger, we have to consider the band shifting, namely the self-energy correction to the electron, denoted as $\delta E_{1,2}(k)$. The gap moves to points where $E_1+\delta E_1+\Omega-E_2-\delta E_2 = 0$ and is compressed by the band shifting. Thus we see that the gaps are shrinking at larger $|\boldsymbol{A}|$. As the field gets even larger such that the resonant points disappear, i.e. $|E_1+\delta E_1-E_2-\delta E_2 |<\Omega$ for all $k$, the gap becomes trivial, which is simply a separation between two shifted Bloch bands. We observe and mark the turning points in Fig.~\ref{fig:FR}(d) which indicates that the gap transfers from a resonant split to a trivial separation. This then explains the decreases in the extrinsic current and the power feature, confirmed by the correspondence of the turning points marked in Fig.~\ref{fig:FR}(b-d). We notice that the intrinsic current decreases a little earlier than the extrinsic, indicating extra possible effects such as more balanced population under stronger field. We emphasize that this strong field effect works in effectively two-band systems and may get complicated in multiband systems. 

\textit{Discussions and Remarks}.--- 
In this work, we discussed DC currents and their power features in a non-interacting Floquet-Bloch system with weak damping and strong drive. The results show the behaviors of the Bloch system under intensive drive. We want to briefly discuss what may happen when considering more aspects.

First is about the weak field limit where the field strength is small enough and $\Gamma\gg\Delta_{\boldsymbol{A}}$. Starting from the density matrix in Eq.~\eqref{FBdensityMatrix}, we can reproduce the well-known non-linear shift current as previously obtained using perturbation theory or Floquet Green's function method~\cite{SM,Morimoto2016Topological,young2012first}. Another important aspect is the electron-electron interactions that has been ignored in this work. We want to note that, technically, the interaction normally leads to heating of the system, which tends to erase notable features like the current responses~\cite{Alessio2014long,Lazarides2014equilibrium,genske2015floquet}. Lastly, the oversimplification of the constant damping rate $\Gamma$ may fail at large field strength. Therefore, the theory needs to further take into account possible field or even momentum dependence of the damping, for example the treatments in Refs.~\cite{dehghani2014dissipative,dehghani2015out}. However, we believe that in some highly controllable platforms, like the cold atom system, the parameters can be tuned to suit the approximations we used in this work. We also claim that despite various damping mechanisms, the formula in Eq.~\eqref{FLe} is general and the intrinsic current response is present under broken TRS, which is important for efficiently generating DC currents.

Q. G. acknowledges financial support of the Provost’s Graduate Excellence Fellowship from The University of Texas at Austin.

\onecolumngrid
\section{Supplementary Material}

\subsection{Primes on Floquet-Bloch eigenbasis}
As discussed in the main text, the Floquet-Bloch Hamiltonian $\hat{H}(\boldsymbol{r},t;\boldsymbol{A})$ can be reduced to a kernel matrix under the extended Floquet-Bloch basis $\{|\phi^\alpha_{n,\boldsymbol{k}}(\boldsymbol{r},t)\rangle \equiv e^{-in\Omega t}|\varphi^\alpha_{\boldsymbol{k}}(\boldsymbol{r})\rangle\}$:
\begin{equation}\label{HamiltonianKernel}
    \begin{split}
    \mathcal{H}_{m,n;\alpha,\beta}(\boldsymbol{k}) &\equiv \langle\langle \phi^\alpha_{m,\boldsymbol{k}}(\boldsymbol{r},t) | \hat{H}(\boldsymbol{r},t;\boldsymbol{A})-i\hbar\partial_t|\phi^\beta_{n,\boldsymbol{k}}(\boldsymbol{r},t)\rangle\rangle .
    \end{split}
\end{equation}
Then we need to solve the eigen-equation: $\sum_{n,\alpha}\mathcal{H}_{m,n;\beta,\alpha}(\boldsymbol{k})f^{\mu,\alpha}_{n,\boldsymbol{k}}=\hbar\omega_\mu(\boldsymbol{k})f^{\mu,\beta}_{n,\boldsymbol{k}}$. We emphasize that the corresponding eigen-energies have a certain pattern: $\{ \omega_{\mu=1}+n\Omega,\cdots, \omega_{\mu=N_B}+n\Omega\}$ with $n$ the Floquet index running over all integers and $N_B$ the number of bands of the corresponding Bloch system. Since the eigen-energies have such periodicity, we can always restrict our considerations within a finite range, say $[0,\Omega]$, which is referred to the Floquet Brillouin zone. The energy confined in such range is called the quasi-energy.

Nevertheless, the kernel matrix $\mathcal{H}_{m,n;\alpha,\beta}(\boldsymbol{k})$ itself defines a complete Hilbert space with eigenvalues as described above and corresponding eigenvectors $(\boldsymbol{f}_{\omega_{\mu}+l\Omega})_{n,\alpha}=f^{\mu,\alpha}_{n+l,\boldsymbol{k}}$. Those eigenvectors form an orthnormal and complete basis, which means that we have
\begin{equation}
    \begin{split}
        \sum_{n,\beta}\left(f_{n,\boldsymbol{k}}^{\mu,\beta}\right)^*f_{n+l,\boldsymbol{k}}^{\nu,\beta} &= \delta_{\mu,\nu}\delta_{l,0}, \\ 
        \sum_{n,\nu}\left(f_{n,\boldsymbol{k}}^{\nu,\alpha}\right)^*f_{n+l,\boldsymbol{k}}^{\nu,\beta} &= \delta_{\alpha,\beta}\delta_{l,0}.
    \end{split}
\end{equation}
Those two relations then enable the Floquet-Bloch eigenbasis $\{ |\tilde{u}^\mu_{\boldsymbol{k}}(\boldsymbol{r},t)\rangle\}$ to be also orthnormal and complete at a specific $\boldsymbol{k}$:
\begin{equation}
    \begin{split}
        \langle\tilde{u}^\mu_{\boldsymbol{k}}(\boldsymbol{r},t)|\tilde{u}^\nu_{\boldsymbol{k}}(\boldsymbol{r},t)\rangle &= \delta_{\mu,\nu}, \\
        \sum_{\mu}|\tilde{u}^\mu_{\boldsymbol{k}}(\boldsymbol{r},t)\rangle\langle\tilde{u}^\mu_{\boldsymbol{k}}(\boldsymbol{r},t)| &= \hat{\boldsymbol{I}}
    \end{split}
\end{equation}
where $ |\tilde{u}^\mu_{\boldsymbol{k}}(\boldsymbol{r},t)\rangle \equiv e^{-i\boldsymbol{k}\cdot\boldsymbol{r}}|\Phi^\mu_{\boldsymbol{k}}(\boldsymbol{r},t)\rangle=\sum_{n,\beta}f^{\mu,\beta}_{n,\boldsymbol{k}} e^{-i n\Omega t}|u^\beta_{\boldsymbol{k}}(\boldsymbol{r},t)\rangle$ is the periodic part of the Floquet-Bloch eigen-wavefunction.

\subsection{Derivation of QLE in FLoquet Picture and the Density Matrix}
The QLE:
\begin{equation}\label{QLE}
    i\partial_t \hat{\rho} (\boldsymbol{k},t) = [\hat{H}(\boldsymbol{k},t), \hat{\rho} (\boldsymbol{k},t)] + i[\hat{D},\hat{\rho}(\boldsymbol{k},t)],
\end{equation}
where
\begin{equation}\label{dampingTerm}
    \begin{split}
    [\hat{D},\hat{\rho}(\boldsymbol{k},t)] = &-\Gamma[\hat{\rho} (\boldsymbol{k},t)-\hat{\rho}^{B,eq} (\boldsymbol{k},t)] \\
    &-\sum_{\alpha\ne \beta}\Gamma'_{\alpha,\beta}(\boldsymbol{k})\rho^B_{\alpha,\beta}(\boldsymbol{k},t)|u^\alpha_{\boldsymbol{k}}\rangle\langle u^\beta_{\boldsymbol{k}}|.
    \end{split}
\end{equation}
% It's easy to write $[\hat{D},\hat{\rho}(\boldsymbol{k},t)]$ in Bloch basis at a specific $\boldsymbol{k}$: \textcolor{blue}{need to be fixed:
% \begin{equation}\label{dampingTerm}
%     \begin{split}
%         &[\hat{D},\hat{\rho}(\boldsymbol{k},t)]_{\alpha,\alpha} = -\Gamma\left[\rho^B_{\alpha,\alpha}(\boldsymbol{k},t)-\rho^{B,eq}_{\alpha,\alpha}(\boldsymbol{k})\right]|u^\alpha_{\boldsymbol{k}}\rangle\langle u^\alpha_{\boldsymbol{k}}| ,\\
%         &[\hat{D},\hat{\rho}(\boldsymbol{k},t)]_{\alpha,\beta} = -\Gamma_{\alpha,\beta}(\boldsymbol{k})\rho^B_{\alpha,\beta}(\boldsymbol{k},t)|u^\alpha_{\boldsymbol{k}}\rangle\langle u^\beta_{\boldsymbol{k}}| ,
%     \end{split}
% \end{equation}
% where $\Gamma_{\alpha,\beta} = \Gamma + \Gamma'_{\alpha,\beta}$.} 
We emphasize here that $\hat{\rho}^{B,eq} (\boldsymbol{k},t)$ is the instantaneous Bloch equilibrium density operator, for which we give a detailed discussion in the next section. Now project the QLE onto the Floquet eigenbasis:
\begin{equation}
    \langle\Tilde{u}^\mu_{\boldsymbol{k}} |i\partial_t \hat{\rho} (\boldsymbol{k},t)|\Tilde{u}^\nu_{\boldsymbol{k}}\rangle = \langle\Tilde{u}^\mu_{\boldsymbol{k}} |[\hat{H}(\boldsymbol{k},t), \hat{\rho} (\boldsymbol{k},t)]|\Tilde{u}^\nu_{\boldsymbol{k}}\rangle + i\langle\Tilde{u}^\mu_{\boldsymbol{k}} |[\hat{D},\hat{\rho}(\boldsymbol{k},t)]|\Tilde{u}^\nu_{\boldsymbol{k}}\rangle,
\end{equation}
where, from the Schrodinger equation, we have 
\begin{equation}
    \begin{split}
    \hat{H}(\boldsymbol{k},t)e^{-i\omega_\mu(\boldsymbol{k})t}|\Tilde{u}^\mu_{\boldsymbol{k}}\rangle &= i\partial_t e^{-i\omega_\mu(\boldsymbol{k})t}|\Tilde{u}^\mu_{\boldsymbol{k}}\rangle = e^{-i\omega_\mu(\boldsymbol{k})t}(\omega_\mu(\boldsymbol{k})+i\partial_t)|\Tilde{u}^\mu_{\boldsymbol{k}}\rangle \\
    &\Rightarrow i\partial_t|\Tilde{u}^\mu_{\boldsymbol{k}}\rangle = (\hat{H}(\boldsymbol{k},t) - \omega_\mu(\boldsymbol{k}))|\Tilde{u}^\mu_{\boldsymbol{k}}\rangle,
    \end{split}
\end{equation}
and then
\begin{equation}
    \begin{split}
        \langle \tilde{u}^\mu_{\boldsymbol{k}} |i\partial_t \hat{\rho} (\boldsymbol{k},t)| \tilde{u}^\nu_{\boldsymbol{k}}\rangle &= i\partial_t \rho^F_{\mu,\nu}(\boldsymbol{k},t)-\langle i\partial_t  \tilde{u}^\mu_{\boldsymbol{k}} |\hat{\rho} (\boldsymbol{k},t)| \tilde{u}^\nu_{\boldsymbol{k}}\rangle-\langle \tilde{u}^\mu_{\boldsymbol{k}} |\hat{\rho} (\boldsymbol{k},t)|i\partial_t  \tilde{u}^\nu_{\boldsymbol{k}}\rangle \\
        &= i\partial_t \rho^F_{\mu,\nu}(\boldsymbol{k},t) + \langle \tilde{u}^\mu_{\boldsymbol{k}} |[\hat{H}(\boldsymbol{k},t), \hat{\rho} (\boldsymbol{k},t)]| \tilde{u}^\nu_{\boldsymbol{k}}\rangle - [\omega_\mu (\boldsymbol{k})-\omega_\nu(\boldsymbol{k})]\rho^F_{\mu,\nu}(\boldsymbol{k},t),
    \end{split}
\end{equation}
where $\rho^F_{\mu,\nu}(\boldsymbol{k},t)\equiv \langle \tilde{u}^\mu_{\boldsymbol{k}} |\hat{\rho} (\boldsymbol{k},t)| \tilde{u}^\nu_{\boldsymbol{k}}\rangle$ is the density matrix element in Floquet eigenbasis. 

Now let's evaluate the damping term:
\begin{equation}
    \begin{split}
        &\langle \tilde{u}^\mu_{\boldsymbol{k}} |[\hat{D}, \hat{\rho} (\boldsymbol{k},t)]| \tilde{u}^\nu_{\boldsymbol{k}}\rangle \\
        =& -\Gamma\langle \tilde{u}^\mu_{\boldsymbol{k}} |\hat{\rho} (\boldsymbol{k},t)| \tilde{u}^\nu_{\boldsymbol{k}}\rangle + \Gamma\langle \tilde{u}^\mu_{\boldsymbol{k}} |\hat{\rho}^{B,eq} (\boldsymbol{k},t)| \tilde{u}^\nu_{\boldsymbol{k}}\rangle -\sum_{\alpha\ne\beta}\Gamma'_{\alpha,\beta}(\boldsymbol{k})\langle \tilde{u}^\mu_{\boldsymbol{k}} |\rho^B_{\alpha,\beta}(\boldsymbol{k},t)| u^\alpha_{\boldsymbol{k}}\rangle\langle u^\beta_{\boldsymbol{k}}| \tilde{u}^\nu_{\boldsymbol{k}}\rangle \\
        =& -\Gamma\rho_{\mu,\nu}^F(\boldsymbol{k},t)+\Gamma\sum_{\alpha,\beta} \rho^{B,eq}_{\alpha,\beta}(\boldsymbol{k},t)\langle \tilde{u}^\mu_{\boldsymbol{k}} | u^\alpha_{\boldsymbol{k}}\rangle\langle u^\beta_{\boldsymbol{k}}| \tilde{u}^\nu_{\boldsymbol{k}}\rangle - \sum_{\alpha\ne\beta}\Gamma'_{\alpha,\beta}(\boldsymbol{k})\langle \tilde{u}^\mu_{\boldsymbol{k}} |\rho^B_{\alpha,\beta}(\boldsymbol{k},t)| u^\alpha_{\boldsymbol{k}}\rangle\langle u^\beta_{\boldsymbol{k}}| \tilde{u}^\nu_{\boldsymbol{k}}\rangle \\
        =& -\Gamma\rho_{\mu,\nu}^F(\boldsymbol{k},t)+\Gamma\Bar{\rho}^{eq}_{\mu,\nu}(\boldsymbol{k})-\mathcal{S}_{\mu,\nu}(\boldsymbol{k},t),
    \end{split}
\end{equation}
where $\Bar{\rho}^{eq}_{\mu,\nu}(\boldsymbol{k})\equiv \sum_{\alpha,\beta} \rho^{B,eq}_{\alpha,\beta}(\boldsymbol{k},t)\langle \tilde{u}^\mu_{\boldsymbol{k}} | u^\alpha_{\boldsymbol{k}}\rangle\langle u^\beta_{\boldsymbol{k}}| \tilde{u}^\nu_{\boldsymbol{k}}\rangle $ is the instantaneous Bloch equilibrium density matrix in Floquet picture and $\mathcal{S}_{\mu,\nu}(\boldsymbol{k},t)\equiv\sum_{\alpha\ne\beta}\Gamma'_{\alpha,\beta}(\boldsymbol{k})\langle \tilde{u}^\mu_{\boldsymbol{k}} |\rho^B_{\alpha,\beta}(\boldsymbol{k})| u^\alpha_{\boldsymbol{k}}\rangle\langle u^\beta_{\boldsymbol{k}}| \tilde{u}^\nu_{\boldsymbol{k}}\rangle$ is the scattering matrix. Putting all together, we have the QLE in Floquet picture:
\begin{equation}\label{QLEinFlouqet}
    i\partial_t\rho^F_{\mu,\nu}(\boldsymbol{k},t) = \left[ \omega_\mu(\boldsymbol{k})- \omega_\nu(\boldsymbol{k})-i\Gamma\right]\rho^F_{\mu,\nu}(\boldsymbol{k},t)+i\Gamma\Bar{\rho}^{eq}_{\mu,\nu}(\boldsymbol{k},t)-i\mathcal{S}_{\mu,\nu}(\boldsymbol{k},t),
\end{equation}

After a long time evolution, the Floquet-Bloch system enters into a stationary state where its changing frequency synchronizes with the Floquet drive frenquecy $\Omega$. In that case, we can do the Fourier transformation:
\begin{equation}
    \begin{split}
        \rho^F_{\mu,\nu}(\boldsymbol{k},t) &= \sum_l \rho^{F,l}_{\mu,\nu}(\boldsymbol{k})e^{-il\Omega t}, \\
        \Bar{\rho}^{eq}_{\mu,\nu}(\boldsymbol{k},t) &= \sum_l \Bar{\rho}^{eq,l}_{\mu,\nu}(\boldsymbol{k})e^{-il\Omega t}, \\
        \mathcal{S}_{\mu,\nu}(\boldsymbol{k},t) &= \sum_l \mathcal{S}^{l}_{\mu,\nu}(\boldsymbol{k})e^{-il\Omega t}.
    \end{split}
\end{equation}
By inserting those Fourier expansions into Eq.~\eqref{QLEinFlouqet}, we can easily obtain a solution for the density matrix:
\begin{equation}\label{densityMatrix}
    \rho^{F,l}_{\mu,\nu}(\boldsymbol{k}) = \frac{-i\Gamma\Bar{\rho}^{eq,l}_{\mu,\nu}(\boldsymbol{k})+i\mathcal{S}^{l}_{\mu,\nu}(\boldsymbol{k})}{\omega_\mu(\boldsymbol{k})-\omega_\nu(\boldsymbol{k})-l\Omega-i\Gamma}.
\end{equation}

\subsection{Instantaneous Bloch equilibrium under static Bloch eigenbasis and its induced dragging effect}

In the QLE, we used the instantaneous Bloch thermal equilibrium $\hat{\rho}^{B,eq}(\boldsymbol{k},t)$ emphasizing the fact that when $\Gamma\to \infty$, the system is always relaxed to the instantaneous Bloch equilibrium with the heat bath, namely the Floquet dynamics is totally washed out by extremely fast damping. Let's first give the proper definition of the instantaneous Bloch eigenstates:
\begin{equation}
    \hat{H}(\boldsymbol{k},\boldsymbol{A}(t))|u^{\alpha}_{\boldsymbol{k}}(t)\rangle = E_{\alpha}(\boldsymbol{k},t)|u^{\alpha}_{\boldsymbol{k}}(t)\rangle
\end{equation}
where $E_{\alpha}(\boldsymbol{k},t)$ and $|u^{\alpha}_{\boldsymbol{k}}(t)\rangle$ are the instantaneous eigenenergy and eigenstate, respectively. One should notice that the time-dependence introduced by the driving field $\boldsymbol{A}(t)$ is simply a parameter and does not enters the dynamics. Then we claim that the instantaneous Bloch equilibrium is diagonal under the instantaneous Bloch basis:
\begin{equation}
    \hat{\rho}^{B,eq}(\boldsymbol{k},t) = \sum_{\alpha}\tilde{\rho}^{B,eq}_{\alpha,\alpha}(\boldsymbol{k},t)|u^{\alpha}_{\boldsymbol{k}}(t)\rangle\langle u^{\alpha}_{\boldsymbol{k}}(t)|.
\end{equation}
We then expand this density matrix in the static Bloch eigenbasis\footnote{This requires that the Hilbert space spanned by the instantaneous Bloch basis $\{|u^{\alpha}_{\boldsymbol{k}}(t)\rangle\}$ is equivalent to that spanned by the static Bloch basis $\{|u^{\alpha}_{\boldsymbol{k}}\rangle\}$, which may not be true. However, one can show that the difference (if exists) can only occur at the second or higher orders in the external field strength.}: 
\begin{equation}\label{ins_den}
    \hat{\rho}^{B,eq}(\boldsymbol{k},t) = \sum_{\alpha,\beta}\rho^{B,eq}_{\alpha,\beta}(\boldsymbol{k},t)|u^{\alpha}_{\boldsymbol{k}}\rangle\langle u^{\beta}_{\boldsymbol{k}}|,
\end{equation}
where the static Bloch eigenbasis corresponds to non-driven Bloch system, i.e. $|u^{\alpha}_{\boldsymbol{k}}\rangle  = |u^{\alpha}_{\boldsymbol{k}}(t)\rangle|_{\boldsymbol{A}(t)=0}$. Now, let's also introduce the static Bloch equilibrium which is the thermal equilibrium without Floquet drive ($\boldsymbol{A}(t)=0$):
\begin{equation}\label{sta_den}
    \hat{\rho}^{B,eq,s}(\boldsymbol{k}) = \sum_{\alpha}\rho^{B,eq,s}_{\alpha,\alpha}(\boldsymbol{k})|u^{\alpha}_{\boldsymbol{k}}\rangle\langle u^{\alpha}_{\boldsymbol{k}}|.
\end{equation}
For example, for Fermi-Dirac distribution, we have $\rho^{B,eq,s}_{\alpha,\alpha}(\boldsymbol{k}) = 1/(e^{[E_{\alpha}(\boldsymbol{k})-\mu_c]/k_B \mathcal{T}}+1)$. 

Then the question is how to determine the instantaneous density matrix in Eq.~\eqref{ins_den} given the static density matrix in Eq.~\eqref{sta_den}. In general, one needs to know the detailed Hamiltonian to answer the question. So, in the following, we specify our consideration to be a coherent light driven system which is also studied in the main text. In such case, the Hamiltonian becomes $ \hat{H}(\boldsymbol{k},\boldsymbol{A}(t))=\hat{H}_B(\boldsymbol{k}+\boldsymbol{A}(t))$ where $\boldsymbol{A}(t)$ is the vector potential. The minimal substitution allows us to easily identify that
\begin{equation}
    \begin{split}
        \tilde{\rho}^{B,eq}_{\alpha,\alpha}(\boldsymbol{k},t) &= \rho^{B,eq,s}_{\alpha,\alpha}(\boldsymbol{k}+\boldsymbol{A}(t)); \\
        |u^{\alpha}_{\boldsymbol{k}}(t)\rangle &= |u^{\alpha}_{\boldsymbol{k}+\boldsymbol{A}(t)}\rangle.
    \end{split}
\end{equation}
Thus, the matrix element $\rho^{B,eq}_{\alpha,\beta}(\boldsymbol{k},t)$ in Eq.~\eqref{ins_den} is given by
\begin{equation}
    \begin{split}
        \rho^{B,eq}_{\alpha,\beta}(\boldsymbol{k},t) &= \langle u^{\alpha}_{\boldsymbol{k}}|\sum_{\eta}\rho^{B,eq,s}_{\eta,\eta}(\boldsymbol{k}+\boldsymbol{A}(t))|u^{\eta}_{\boldsymbol{k}+\boldsymbol{A}(t)}\rangle\langle u^{\eta}_{\boldsymbol{k}+\boldsymbol{A}(t)}|u^{\beta}_{\boldsymbol{k}}\rangle \\
        &= \sum_{\eta}\left[ \rho^{B,eq,s}_{\eta,\eta}(\boldsymbol{k}) + \boldsymbol{A}(t)\cdot\frac{\partial}{\partial\boldsymbol{k}}\rho^{B,eq,s}_{\eta,\eta}(\boldsymbol{k})+\cdots\right]\langle u^{\alpha}_{\boldsymbol{k}}|\left[|u^{\eta}_{\boldsymbol{k}}\rangle + \boldsymbol{A}(t)\cdot\frac{\partial}{\partial\boldsymbol{k}}|u^{\eta}_{\boldsymbol{k}}\rangle + \cdots\right] \\
        &\qquad\times\left[\langle u^{\eta}_{\boldsymbol{k}}| + \boldsymbol{A}(t)\cdot\frac{\partial}{\partial\boldsymbol{k}}\langle u^{\eta}_{\boldsymbol{k}}| + \cdots\right]|u^{\beta}_{\boldsymbol{k}}\rangle \\
        &= \sum_{\eta}\left[ \rho^{B,eq,s}_{\eta,\eta}(\boldsymbol{k}) + \boldsymbol{A}(t)\cdot\frac{\partial}{\partial\boldsymbol{k}}\rho^{B,eq,s}_{\eta,\eta}(\boldsymbol{k})+\cdots\right]\left(\delta_{\alpha,\eta}-i\boldsymbol{A}(t)\cdot\boldsymbol{\mathcal{A}}^{B}_{\alpha,\eta}(\boldsymbol{k})+\cdots \right) \\
        &\qquad\times\left(\delta_{\eta,\beta}+i\boldsymbol{A}(t)\cdot\boldsymbol{\mathcal{A}}^{B}_{\eta,\beta}(\boldsymbol{k})+\cdots \right),
    \end{split}
\end{equation}
where $\boldsymbol{\mathcal{A}}^{B}_{\alpha,\beta}(\boldsymbol{k})\equiv \langle u^{\alpha}_{\boldsymbol{k}}|i\partial_{\boldsymbol{k}}|u^{\beta}_{\boldsymbol{k}}\rangle$ is the interband Bloch Berry connection. However, the expression is still very complicated, which needs to be simplified. Recall that $\rho^{B,eq,s}_{\eta,\eta}(\boldsymbol{k})$ is the static Bloch equilibrium distribution, we have that for low temperature, if the corresponding Bloch system obeys Fermi-Dirac distribution and the chemical potential is in the gap, we can ignore the derivatives of $\rho^{B,eq,s}_{\eta,\eta}(\boldsymbol{k})$ in the above equation. One should anticipate that there could be large corrections if the temperature is high or the chemical potential is not in the gap. The matrix element can be written up to the second order in the field strength $|\boldsymbol{A}(t)|$ as:
\begin{equation}\label{ins_den_element}
    \begin{split}
        \rho^{B,eq}_{\alpha,\beta}(\boldsymbol{k},t) &= \rho^{B,eq,s}_{\alpha,\alpha}(\boldsymbol{k})\delta_{\alpha,\beta} + i\left[\boldsymbol{A}(t)\cdot\boldsymbol{\mathcal{A}}^{B}_{\alpha,\beta}(\boldsymbol{k})+\frac{1}{2}\boldsymbol{A}(t)\boldsymbol{A}(t)\cdot\frac{\partial}{\partial\boldsymbol{k}}\boldsymbol{\mathcal{A}}^{B}_{\alpha,\beta}(\boldsymbol{k})\right](\rho^{B,eq,s}_{\alpha,\alpha}-\rho^{B,eq,s}_{\beta,\beta})\\
        &+ \boldsymbol{A}(t)\boldsymbol{A}(t)\cdot\sum_{\eta}\boldsymbol{\mathcal{A}}^{B}_{\alpha,\eta}(\boldsymbol{k})\boldsymbol{\mathcal{A}}^{B}_{\eta,\beta}(\boldsymbol{k})\left(\rho^{B,eq,s}_{\eta,\eta}- \frac{\rho^{B,eq,s}_{\alpha,\alpha}+\rho^{B,eq,s}_{\beta,\beta}}{2} \right) + O(|\boldsymbol{A}(t)|^3),
    \end{split}
\end{equation}
where we can see the first term on the right hand side is exactly the static Bloch equilibrium, while the rest field-dependent terms are corrections that we define as the field dragging effect. To further simplify our consideration, we only showcase the first order dragging correction in the case study in the main text, which we approximate:
\begin{equation}\label{dragging}
    \rho^{B,eq}_{\alpha,\beta}(\boldsymbol{k},t) \approx \rho^{B,eq,s}_{\alpha,\alpha}\delta_{\alpha,\beta} + i\boldsymbol{A}(t)\cdot\boldsymbol{\mathcal{A}}^{B}_{\alpha,\beta}(\boldsymbol{k})(\rho^{B,eq,s}_{\alpha,\alpha}-\rho^{B,eq,s}_{\beta,\beta}),
\end{equation}
because we believe that the higher order terms give only quantitative differences that won't change the physics in the simple two-band system we investigated. Another interesting thing is that the instantaneous density matrix in Eq.~\eqref{ins_den_element} is actually gauge-dependent since it is related to the Berry connections. However, we find that the gauge dependence is exactly canceled out by the corresponding gauge dependence of the Floquet coefficient $f_{n,\boldsymbol{k}}^{\mu,\alpha}$. One can then check the gauge invariance of the density matrix in Eq. (9) in the main text.

\subsection{Derivation of The Floquet-Liouville equation}
In this section we will derive the Floquet-Liouville equation as stated in the main text that for an arbitrary operator $\hat{\mathcal{O}}$ that is time-independent or periodic in time with Floquet period, we have
\begin{equation}\label{FLe}
    \text{Tr}^F[\hat{\rho}\partial_t\hat{\mathcal{O}}]=-i\text{Tr}^F[\hat{\rho}[\hat{H}-\omega,\hat{\mathcal{O}}]]-\text{Tr}^F[[\hat{D},\hat{\rho}]\hat{\mathcal{O}}],
\end{equation}
where $\text{Tr}^F[\sbullet]\equiv\sum_{\mu}\langle\langle\Tilde{u}^\mu_{\boldsymbol{k}} |\sbullet|\Tilde{u}^\mu_{\boldsymbol{k}}\rangle\rangle$ is the Floquet spacetime trace and $\omega$ is the quasi-energy of corresponding Floquet eigenstate: $\omega|\Tilde{u}^\mu_{\boldsymbol{k}}\rangle\rightarrow\omega_\mu(\boldsymbol{k})|\Tilde{u}^\mu_{\boldsymbol{k}}\rangle$. The derivation is quite straightforward. We consider the following quantity:
\begin{equation}\label{FLe_derivation}
    \begin{split}
        \frac{d}{dt}\langle\Tilde{u}^\mu_{\boldsymbol{k}} |\hat{\rho}\hat{\mathcal{O}}|\Tilde{u}^\mu_{\boldsymbol{k}}\rangle &= \langle\partial_t \Tilde{u}^\mu_{\boldsymbol{k}} |\hat{\rho}\hat{\mathcal{O}}|\Tilde{u}^\mu_{\boldsymbol{k}}\rangle + \langle\Tilde{u}^\mu_{\boldsymbol{k}} |\partial_t(\hat{\rho}\hat{\mathcal{O}})|\Tilde{u}^\mu_{\boldsymbol{k}}\rangle + \langle\Tilde{u}^\mu_{\boldsymbol{k}} |\hat{\rho}\hat{\mathcal{O}}|\partial_t\Tilde{u}^\mu_{\boldsymbol{k}}\rangle \\
        &= i\langle\Tilde{u}^\mu_{\boldsymbol{k}} |[\hat{H}-\omega,\hat{\rho}\hat{\mathcal{O}}]|\Tilde{u}^\mu_{\boldsymbol{k}}\rangle + \langle\Tilde{u}^\mu_{\boldsymbol{k}} |(\partial_t\hat{\rho})\hat{\mathcal{O}}|\Tilde{u}^\mu_{\boldsymbol{k}}\rangle + \langle\Tilde{u}^\mu_{\boldsymbol{k}} |\hat{\rho}\partial_t\hat{\mathcal{O}}|\Tilde{u}^\mu_{\boldsymbol{k}}\rangle.
    \end{split}
\end{equation}
Then, recall the quantum Liouville equation and the fact that $\frac{1}{T}\int_0^T \frac{d}{dt}\langle\Tilde{u}^\mu_{\boldsymbol{k}} |\hat{\rho}\hat{\mathcal{O}}|\Tilde{u}^\mu_{\boldsymbol{k}}\rangle = \langle\Tilde{u}^\mu_{\boldsymbol{k}} |\hat{\rho}\hat{\mathcal{O}}|\Tilde{u}^\mu_{\boldsymbol{k}}\rangle|_{t=T}-\langle\Tilde{u}^\mu_{\boldsymbol{k}} |\hat{\rho}\hat{\mathcal{O}}|\Tilde{u}^\mu_{\boldsymbol{k}}\rangle|_{t=0} = 0$, Eq.~\eqref{FLe_derivation}
ends up to be Eq.~\eqref{FLe} after taking the time average and summing over index $\mu$.

One thing we should notice is that the quasi-energy $\omega$ is not just a number in the commutator, instead it has dependence on momentum $\boldsymbol{k}$. This then renders the facts that for density matrix $\hat{\rho}$, we have $[\hat{\rho},\omega] = 0$; but for operator like $i\partial_{\boldsymbol{k}}$, we have $[i\partial_{\boldsymbol{k}},\omega] = i\partial_{\boldsymbol{k}}\omega(\boldsymbol{k})$.

\subsection{DC current response and Floquet power input}
Now, in favor of the Floquet-Liouville equation derived in last section, we can obtain some DC responses. For current response, we choose $\hat{\mathcal{O}} = i\partial_{\boldsymbol{k}}$ in Eq.~\eqref{FLe}, and obtain the following result:
\begin{equation}
    \text{Tr}^F[\hat{\rho}\hat{\boldsymbol{v}}] = \sum_{\mu}\rho^{F,l=0}_{\mu,\mu}\frac{\partial \omega_{\mu}(\boldsymbol{k})}{\partial \boldsymbol{k}} - \text{Tr}^{F}[[\hat{D},\hat{\rho}]i\partial_{\boldsymbol{k}}],
\end{equation}
where $\hat{\boldsymbol{v}}=\frac{\partial\hat{H}(\boldsymbol{k},t)}{\partial \boldsymbol{k}}$ is the velocity operator. Thus the DC current can be calculated using $\boldsymbol{j}_{\text{DC}} = -e\int d\boldsymbol{k}\text{Tr}^F[\hat{\rho}\hat{\boldsymbol{v}}]$. In the main text, we ignored the electron-electron interaction which is to set $\mathcal{S}^{l}_{\mu,\nu}(\boldsymbol{k}) = 0$ in Eq.~\eqref{densityMatrix}. In such setting, $\rho^{F,l=0}_{\mu,\mu} = (\hat{\rho}^{F,eq})_{\mu,\mu}$ is just the Floquet-Bloch quasi-equilibrium distribution and the damping term is reduced to $[\hat{D},\hat{\rho}]=-\Gamma(\hat{\rho}-\hat{\rho}^{B,eq}) $. At the weak damping and strongly driven regime, i.e. $\Gamma \ll \Delta_{\boldsymbol{A}}\equiv\min_{\mu\ne\nu,l}\{ |\omega_\mu(\boldsymbol{k})-\omega_\nu(\boldsymbol{k})-l\Omega|\}$, we also have $ \hat{\rho} = \hat{\rho}^{F,eq} + O(\frac{\Gamma}{\Delta_{\boldsymbol{A}}-i\Gamma})$. Putting all together, we have the result in the main text that:
\begin{equation}\label{velocityEV}
    \text{Tr}^F[\hat{\rho}\hat{\boldsymbol{v}}] = \text{Tr}^F[\hat{\rho}^{F,eq}\hat{\boldsymbol{v}}] +  \Gamma\times\text{Tr}^F[(\hat{\rho}^{F,eq}-\hat{\rho}^{B,eq})i\partial_{\boldsymbol{k}}]+O(\frac{\Gamma^2}{\Delta_{\boldsymbol{A}}-i\Gamma}).
\end{equation}
One can also check the above expression by explicitly evaluating the spacetime trace $\text{Tr}^F[\hat{\rho}\hat{\boldsymbol{v}}]$.

For Floquet power input, we choose $\hat{\mathcal{O}} = \hat{H}$ in Eq.~\eqref{FLe}, and also adopt the above used approximations, ending up with
\begin{equation}\label{power_input}
     \text{Tr}^F[\hat{\rho}\partial_t\hat{H}]= \Gamma\times\text{Tr}^F[(\hat{\rho}^{F,eq}-\hat{\rho}^{B,eq})\hat{H}]+O(\frac{\Gamma^2}{\Delta_{\boldsymbol{A}}-i\Gamma}).
\end{equation}
To illustrate that $\text{Tr}^F[\hat{\rho}\partial_t\hat{H}]$ gives the average energy that the Floquet drive inputs into the system, we now specify the drive as coherent light with vector potential $\boldsymbol{A}(t)$. Thus the Hamiltonian takes the form $ \hat{H}(\boldsymbol{k},t)=\hat{H}_B(\boldsymbol{k}-q\boldsymbol{A}(t))$, where $\hat{H}_B$ is the Hamiltonian of the static Bloch system and $q=-e$ is the charge. Then, we have
\begin{equation}
    \frac{\partial\hat{H}}{\partial t} = \frac{\partial q\boldsymbol{A}}{\partial t}\cdot\frac{\partial\hat{H}}{\partial q\boldsymbol{A}} = -q\frac{\partial\boldsymbol{A}}{\partial t}\cdot\frac{\partial\hat{H}}{\partial \boldsymbol{k}} = q\boldsymbol{E}\cdot\hat{\boldsymbol{v}},
\end{equation}
where $\boldsymbol{E}=-\partial_t\boldsymbol{A}$ is the external electric field. Therefore, the Floquet power input averaged over one Floquet period is just $\Bar{P}(\boldsymbol{k}) \equiv\text{Tr}^{F}[q\boldsymbol{E}(t)\cdot\hat{\rho}\hat{\boldsymbol{v}}]=\text{Tr}^F[\hat{\rho}\partial_t\hat{H}]$. We argue that for other types of Floquet derive like phononic drive, the power input should also have the general form in Eq.~\eqref{power_input}.

\subsection{Gauge Dependence of the DC current}
Let's talk about the gauge dependence of the DC current $\boldsymbol{j}$. As a physical observable, it should be gauge invariant. We are only interested in two major contributions in Eq.~\eqref{velocityEV}: intrinsic current at zeroth order in $\Gamma$; and non-equilibrium current at first order in $\Gamma$. The intrinsic contribution is manifestly gauge invariant since it only involves the equilibrium Floquet-Bloch density matrix, while the second contribution requires a little analysis to check whether it is gauge invariant. 

The primary gauge dependence we consider here comes from the $U(1)$ gauge of the wavefunction:
\begin{equation}
    | \tilde{u}^{\mu}_{\boldsymbol{k}}\rangle \rightarrow e^{-i \theta^{\mu}_{\boldsymbol{k}}} | \tilde{u}^{\mu}_{\boldsymbol{k}}\rangle.
\end{equation}
Upon this gauge transformation, the velocity expectation value will have an extra term
\begin{equation}
    \Gamma\times \text{Tr}^F[(\hat{\rho}^{F,eq}-\hat{\rho}^{B,eq})\partial_{\boldsymbol{k}}\boldsymbol{ \theta}],
\end{equation}
which is zero because
\begin{equation}
    \begin{split}
         \text{Tr}^F[(\hat{\rho}^{F,eq}-\hat{\rho}^{B,eq})\partial_{\boldsymbol{k}}\boldsymbol{ \theta}]=& \sum_{n,\mu,\beta}\rho^{F,0}_{\mu,\mu}(\boldsymbol{k})|f^{\mu,\beta}_{n,\boldsymbol{k}}|^2\partial_{\boldsymbol{k}} \theta^{\mu}_{\boldsymbol{k}}-\sum_{\mu}\langle\langle\tilde{u}^\mu_{\boldsymbol{k}}|\hat{\rho}^{B,eq}|\tilde{u}^\mu_{\boldsymbol{k}}\rangle\rangle\partial_{\boldsymbol{k}} \theta^{\mu}_{\boldsymbol{k}}\\
        =& \sum_{\mu}\rho^{F,0}_{\mu,\mu}(\boldsymbol{k})\left(\sum_{n,\beta}|f^{\mu,\beta}_{n,\boldsymbol{k}}|^2\right)\partial_{\boldsymbol{k}} \theta^{\mu}_{\boldsymbol{k}} - \sum_{\mu}\rho^{F,0}_{\mu,\mu}(\boldsymbol{k})\partial_{\boldsymbol{k}} \theta^{\mu}_{\boldsymbol{k}} \\
        =& \sum_{\mu}\left(\rho^{F,0}_{\mu,\mu}(\boldsymbol{k})-\rho^{F,0}_{\mu,\mu}(\boldsymbol{k})\right)\partial_{\boldsymbol{k}} \theta^{\mu}_{\boldsymbol{k}} = 0.
    \end{split}
\end{equation}
% \begin{equation}
%     \begin{split}
%         & \text{Tr}^F[(\hat{\rho}^{F,eq}-\hat{\rho}^{B,eq})\partial_{\boldsymbol{k}}\boldsymbol{ \theta}] \\
%         =& \sum_{n,\mu,\beta}\left(\rho^{F,0}_{\mu,\mu}(\boldsymbol{k})- \rho^{B,eq}_{\beta,\beta}(\boldsymbol{k}) \right)|f^{\mu,\beta}_{n,\boldsymbol{k}}|^2\partial_{\boldsymbol{k}} \theta^{\mu}_{\boldsymbol{k}} \\
%         =& \sum_{\mu}\rho^{F,0}_{\mu,\mu}(\boldsymbol{k})\left(\sum_{n,\beta}|f^{\mu,\beta}_{n,\boldsymbol{k}}|^2\right)\partial_{\boldsymbol{k}} \theta^{\mu}_{\boldsymbol{k}} - \sum_{\mu}\left(\sum_{n,\beta} \rho^{B,eq}_{\beta,\beta}(\boldsymbol{k})|f^{\mu,\beta}_{n,\boldsymbol{k}}|^2 \right)\partial_{\boldsymbol{k}} \theta^{\mu}_{\boldsymbol{k}} \\
%         =& \sum_{\mu}\left(\rho^{F,0}_{\mu,\mu}(\boldsymbol{k})-\rho^{F,0}_{\mu,\mu}(\boldsymbol{k})\right)\partial_{\boldsymbol{k}} \theta^{\mu}_{\boldsymbol{k}} = 0.
%     \end{split}
% \end{equation}
So, the current is gauge invariant. We can then fix the gauge and recast the expression of the current into a totally diagonal form. By choosing the following special gauge:
\begin{equation}\label{gaugeFixing}
    \partial_{\boldsymbol{k}} \theta^{\mu}_{\boldsymbol{k}} = \frac{\langle\langle \tilde{u}^\mu_{\boldsymbol{k}} |\hat{\rho}^{B,eq}i\partial_{\boldsymbol{k}}| \tilde{u}^\mu_{\boldsymbol{k}}\rangle\rangle}{\rho^{F,0}_{\mu,\mu}(\boldsymbol{k})},
\end{equation}
we have the final form of the expectation value of velocity operator as
\begin{equation}
    \text{Tr}^F[\hat{\rho}\hat{\boldsymbol{v}}] = \text{Tr}^F[\hat{\rho}^{F,eq}\Tilde{\hat{\boldsymbol{v}}}]
\end{equation}
where $\Tilde{\hat{\boldsymbol{v}}}(\boldsymbol{k},t;\Gamma) \equiv \hat{\boldsymbol{v}}+\Gamma(i\partial_{\boldsymbol{k}}+ \text{gauge term})$, defined as the shifted velocity operator. We have to note here that the operator $\Tilde{\hat{\boldsymbol{v}}}$ is in general gauge dependent, but the gauge term vanishes at condition in Eq.~\eqref{gaugeFixing}.

\subsection{Reproducing the non-linear optical Shift Current in the weak field limit}
In previous sections, we derived a formula for current response using the density matrix approach and discussed the result at the large field limit which is $\Gamma \ll \Delta_{\boldsymbol{A}}$. In this section, we want to show that at weak field limit where $\Gamma \gg \Delta_{\boldsymbol{A}}$, the density matrix formalism can reproduce the result of the well-known shift current if the time reversal symmetry (TRS) is preserved.

The main idea is that with small field strength $|\boldsymbol{A}|$, we can truncate the Hamiltonian kernel matrix $\mathcal{H}_{m,n;\alpha,\beta}(\boldsymbol{k})$ to a minimum $2\times 2$ matrix~\cite{Morimoto2016Topological}:
\begin{equation}\label{truncatedKernel}
		\mathcal{H}_{2\times 2}(\boldsymbol{k}) =  \begin{bmatrix}
				E_1(k)+\hbar\Omega & e\boldsymbol{A}^*\cdot\boldsymbol{v}^B_{12}  \\
	        	e\boldsymbol{A}\cdot\boldsymbol{v}^B_{21} & E_2(k)
			\end{bmatrix},
\end{equation}
where $\boldsymbol{v}^B_{\alpha\beta}(\boldsymbol{k})=\langle u^\alpha_{\boldsymbol{k}}|\partial_{\boldsymbol{k}}H_B(\boldsymbol{k})|u^\beta_{\boldsymbol{k}}\rangle$ is the velocity matrix of the original Bloch system. Now, the model is simple enough so that we can have an analytical expression for the DC current in non-interacting Floquet-Bloch systems where again $\mathcal{S}^{l}_{\mu,\nu}(\boldsymbol{k}) = 0$ in Eq.~\eqref{densityMatrix}. For simplicity, we also ignore the field dragging effect described in Eq.~\eqref{dragging} at small field strength $|\boldsymbol{A}|$. The DC current can then be fully evaluated without assuming the scale of $\Gamma$ and the result reads
\begin{equation}\label{finalCurrent}
    \begin{split}
        \boldsymbol{j}_{\text{DC}} =& -e\int \text{d}\boldsymbol{k}\left[(\rho^{B,eq}_{1,1}+\rho^{B,eq}_{2,2}) \frac{\partial_{\boldsymbol{k}}[E_1(\boldsymbol{k})+E_2(\boldsymbol{k})]}{2\hbar} \right. \\
        &+\left. (\rho^{B,eq}_{1,1}-\rho^{B,eq}_{2,2})\left(\frac{\hbar\Omega-\Delta^B_{\boldsymbol{k}}}{2\hbar}\partial_{\boldsymbol{k}}\log\Delta_{\boldsymbol{k}}+\frac{2\Gamma|e\boldsymbol{A}^*\cdot\boldsymbol{v}^B_{12}|^2}{\Delta_{\boldsymbol{k}}^2+\Gamma^2}\boldsymbol{\mathcal{R}}_{\boldsymbol{k}}+\frac{2\Gamma^2|e\boldsymbol{A}^*\cdot\boldsymbol{v}^B_{12}|}{\Delta_{\boldsymbol{k}}^2+\Gamma^2}\boldsymbol{\chi}_{\boldsymbol{k}}\right)\right],
    \end{split}
\end{equation}
where $\Delta^B_{\boldsymbol{k}}\equiv E_2(\boldsymbol{k})-E_1(\boldsymbol{k})$, $\Delta_{\boldsymbol{k}}\equiv \sqrt{(E_1+\hbar\Omega-E_2)^2+4|e\boldsymbol{A}^*\cdot\boldsymbol{v}^B_{12}|^2}$ are the gap functions for the original Bloch system and the Floquet-Bloch system, respectively. Therefore, the minimum direct gap defined before is just $\Delta_{\boldsymbol{A}} = 2|e\boldsymbol{A}^*\cdot\boldsymbol{v}^B_{12}|$. $\boldsymbol{\mathcal{R}}_{\boldsymbol{k}}\equiv\partial_{\boldsymbol{k}}\arg(\boldsymbol{v}^B_{12}) + \boldsymbol{\mathcal{A}}^B_2-\boldsymbol{\mathcal{A}}^B_1$ is the so-called shift vector, with $\boldsymbol{\mathcal{A}}^B_\alpha=\langle u^\alpha_{\boldsymbol{k}}|i\partial_{\boldsymbol{k}}|u^\alpha_{\boldsymbol{k}}\rangle$ being the Bloch Berry connection. $\boldsymbol{\chi}_{\boldsymbol{k}}$ denotes $[|e\boldsymbol{A}^*\cdot\boldsymbol{v}^B_{12}|\partial_{\boldsymbol{k}}(E_1+\hbar\Omega-E_2)-(E_1+\hbar\Omega-E_2)\partial_{\boldsymbol{k}}|e\boldsymbol{A}^*\cdot\boldsymbol{v}^B_{12}|]/\Delta_{\boldsymbol{k}}^2$ for convenience. We note that the term involving $\boldsymbol{\chi}_{\boldsymbol{k}}$ scales roughly as $1/|\boldsymbol{A}|$, which makes it important in weak field, while it is not so important in strong field regime.
Now, if we assume the system has TRS, then most of the contributions in Eq.~\eqref{finalCurrent} vanish because that the corresponding integrands are odd under $\boldsymbol{k}\to - \boldsymbol{k}$. The current is further reduced to
\begin{equation}\label{TRS_DC}
    \boldsymbol{j}^{\text{TRS}}_{\text{DC}} = -e\int \text{d}\boldsymbol{k}(\rho^{B,eq}_{1,1}-\rho^{B,eq}_{2,2})\frac{2\Gamma|e\boldsymbol{A}^*\cdot\boldsymbol{v}^B_{12}|^2}{\Delta_{\boldsymbol{k}}^2+\Gamma^2}\boldsymbol{\mathcal{R}}_{\boldsymbol{k}}.
\end{equation}
We adopt the idea used in Ref.~\cite{Morimoto2016Topological} that in the limit where $\Gamma$ and $|e\boldsymbol{A}^*\cdot\boldsymbol{v}^B_{12}|$ are much smaller than the energy dispersion and $\Gamma\gg|e\boldsymbol{A}^*\cdot\boldsymbol{v}^B_{12}|$, we have 
\begin{equation}
    \boldsymbol{j}^{\text{TRS}}_{\text{DC}} = -e\frac{2\pi|e\boldsymbol{E}|^2}{\Omega^2}\int \text{d}\boldsymbol{k}(\rho^{B,eq}_{1,1}-\rho^{B,eq}_{2,2})\delta(E_1+\hbar\Omega-E_2)|\boldsymbol{v}^B_{12}|^2\boldsymbol{\mathcal{R}}_{\boldsymbol{k}},
\end{equation}
which is exactly the non-linear optical shift current~\cite{young2012first}. Here $|\boldsymbol{E}|=\Omega|\boldsymbol{A}|$ is the corresponding electric field strength.

\subsection{Power signature and its implication on Floquet dynamics}
One can follow the same procedure in the previous section and obtain the power input in a relatively weak field regime as:
\begin{equation}
    \Bar{P}_{\text{net}}= \int \text{d}\boldsymbol{k}(\rho^{B,eq}_{1,1}-\rho^{B,eq}_{2,2})\frac{2\Gamma\Omega|e\boldsymbol{A}^*\cdot\boldsymbol{v}^B_{12}|^2}{\Delta_{\boldsymbol{k}}^2+\Gamma^2}.
\end{equation}
We want to simplify our discussion by excluding the contribution from terms involving $\boldsymbol{\chi}_{\boldsymbol{k}}$ in Eq.~\eqref{finalCurrent} as it becomes less important in the strong field regime. However, we notice it could give a large correction in the weak field and we leave it to a future study. An observation of the net power is that this power input is always positive as long as the population difference between two bands is positive. 

\begin{figure}
    \centering
    \includegraphics[width=6.5cm]{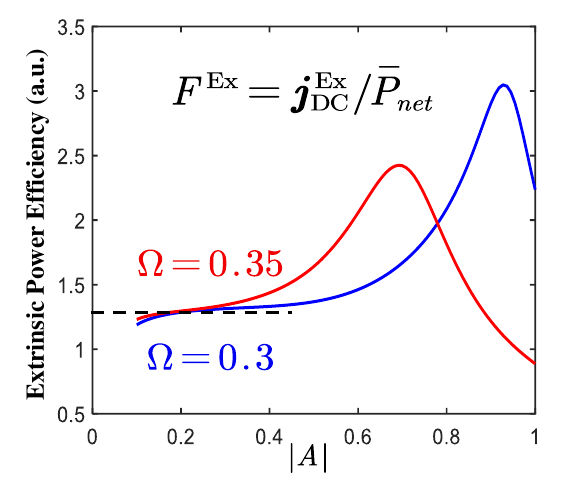}
    \caption{The extrinsic power efficiency $F^{\text{Ex}} $ as functions of $|\boldsymbol{A}|$ for $\Omega=0.3$ (blue) and $\Omega=0.35$ (red), respectively. Parameters used are the same as in the main text.}
    \label{fig:ExPE}
\end{figure}

Let's now look at the power efficiency for generating the extrinsic current which has an analytic expression in the weak field regime:
\begin{equation}
    F^{\text{Ex}} = \frac{\boldsymbol{j}^{\text{Ex}}_{\text{DC}}}{\Bar{P}_{\text{net}}} = \frac{\int \text{d}\boldsymbol{k}(\rho^{B,eq}_{1,1}-\rho^{B,eq}_{2,2})\frac{2\Gamma|\boldsymbol{A}^*\cdot\boldsymbol{v}^B_{12}|^2}{\Delta_{\boldsymbol{k}}^2+\Gamma^2}\boldsymbol{\mathcal{R}}_{\boldsymbol{k}}}{\int \text{d}\boldsymbol{k}(\rho^{B,eq}_{1,1}-\rho^{B,eq}_{2,2})\frac{2\Gamma\Omega|\boldsymbol{A}^*\cdot\boldsymbol{v}^B_{12}|^2}{\Delta_{\boldsymbol{k}}^2+\Gamma^2}}\equiv\frac{\langle\boldsymbol{\mathcal{R}}_{\boldsymbol{k}}\rangle_{\boldsymbol{A}}}{\Omega},
\end{equation}
where we have omitted the electric charge $(-e)$ and $\langle\boldsymbol{\mathcal{R}}_{\boldsymbol{k}}\rangle_{\boldsymbol{A}} $ is defined as the dynamical average of the shift vector $\boldsymbol{\mathcal{R}}_{\boldsymbol{k}}$ in the transition process. If we assume a fully filled Bloch band $E_1$ and an empty $E_2$, then in the weak field and sufficient small damping limit, the extrinsic power efficiency becomes
\begin{equation}
    F^{\text{Ex}} \approx \frac{\sum_{\boldsymbol{k}_i}\boldsymbol{\mathcal{R}}_{\boldsymbol{k}_i}|\frac{\boldsymbol{v}^B_{12}(\boldsymbol{k}_i)}{\boldsymbol{v}^B_{11}(\boldsymbol{k}_i)-\boldsymbol{v}^B_{22}(\boldsymbol{k}_i)}|}{\Omega\sum_{\boldsymbol{k}_i}|\frac{\boldsymbol{v}^B_{12}(\boldsymbol{k}_i)}{\boldsymbol{v}^B_{11}(\boldsymbol{k}_i)-\boldsymbol{v}^B_{22}(\boldsymbol{k}_i)}|},
\end{equation}
which is field-independent. The summation is over all resonant $\boldsymbol{k}_i$ such that $E_1(\boldsymbol{k}_i)+\hbar\Omega-E_2(\boldsymbol{k}_i) = 0$. For the two-band model considered in the main text, we can estimate this factor to be $F^{\text{Ex}}\approx 1.26 $ for $\Omega = 0.3$ and $F^{\text{Ex}}\approx 1.24 $ for $\Omega = 0.35$, respectively. Although this field-independent feature is found in the weak field limit, in the Fig.~\ref{fig:ExPE} where we plot $F^{\text{Ex}}$ as a function of field strength $|\boldsymbol{A}|$, we do observe constant-like behaviors in the range of relatively small $|\boldsymbol{A}|$ and values for two different $\Omega$'s are closed to the values estimated above. However, the efficiency shows strong field dependence when the field is getting stronger. Thus, we say that the extrinsic power efficiency is of particular importance that reflects the Floquet dynamics.


\begin{thebibliography}{35}
\bibitem{Notations}
    In this work, we use the following convention for notations: $(\alpha,\beta)$ and $(\mu,\nu)$ are band indies for Bloch systems and Floquet-Bloch systems, respectively; $(m,n)$ are the multiplicities of the Floquet frequency $\Omega$ in the Floquet exponents; $|\Phi^\mu_{\boldsymbol{k}}\rangle=e^{i\boldsymbol{k}\cdot\boldsymbol{r}}|\Tilde{u}^\mu_{\boldsymbol{k}}\rangle$ and $|\varphi^\beta_{\boldsymbol{k}}\rangle=e^{i\boldsymbol{k}\cdot\boldsymbol{r}}|u^\beta_{\boldsymbol{k}}\rangle$ stand for Floquet-Bloch state and Bloch state, respectively. An example will be the Floquet-Bloch eigenstate: $|\Phi^\mu_{\boldsymbol{k}}(\boldsymbol{r},t)\rangle  \equiv \sum_{n,\beta}f^{\mu,\beta}_{n,\boldsymbol{k}}e^{-in\Omega t}|\varphi^\beta_{\boldsymbol{k}}(\boldsymbol{r},t)\rangle$, where the Floquet exponent multiplicity $n$ and the Bloch band index $\beta$ are summed over.
\bibitem{BlochBandParameters}
	The parameters are the following: $\varepsilon_1$, $\varepsilon_2$ are two on-site energies of two atoms in single unit cell; $d$, $s$ are the lattice constant and distance between two atoms within a unit cell, respectively; $t_1$, $t_2$ and $t_3$, $t_4$ are the nearest neighbor and next nearest neighbor hopping constants, respectively.
\end{thebibliography}
\end{document}